\documentclass[a4paper,12pt]{article}

\usepackage{jheppub} 

\usepackage[T1]{fontenc} 
\usepackage{latexsym,amsmath,amsfonts,amssymb,amsthm}
\usepackage{amsmath}
\usepackage{mathtools}
\usepackage{xcolor}
\usepackage{setspace}

\title{Flavored modular differential equations}

\author{Yiwen Pan,}
\author{Yufan Wang}

\affiliation{School of Physics, Sun Yat-sen University,\\No. 135 Xingangxi Road, Guangzhou, Guangdong, China}

\emailAdd{panyw5@mail.sysu.edu.cn}
\emailAdd{wangyf228@mail2.sysu.edu.cn}

\abstract{
  Flavored modular differential equations sometimes arise from null states or their descendants in a chiral algebra with continuous flavor symmetry. In this paper we focus on Kac-Moody algebras $\widehat{\mathfrak{g}}_k$ that contain a level-four null state $|\mathcal{N}_T\rangle$ which implements the nilpotency of the Sugawara stress tensor. We study the properties of the corresponding flavored modular differential equations, and show that the equations exhibit almost covariance under modular $S$-transformation, connecting null states and their descendants at different levels. The modular property of the equations fixes the structure of $\mathfrak{g}$ and the level $k$, as well as the flavored characters of all the highest weight representations. Shift property of the equations can generate non-vacuum characters starting from the vacuum character.
}

\begin{document} 
\maketitle
\flushbottom


\section{Introduction}

Conformal field theory (CFT) is one of the central players in physics and mathematics. In two dimensions, CFTs are strongly constrained by the infinite dimensional chiral symmetry algebras and the modular group, such that a classification of 2d CFTs is an important and also a likely approachable problem. Among all 2d CFTs, the rational conformal field theories (RCFTs) are the simplest from the classification point of view. One crucial tool to this task is the null states\footnote{Also refereed to as null vectors, singular vectors or just nulls in the literature.}, from which differential equations of correlation functions on $\mathbb{C}$ can be derived and used to constrain the correlation functions and the spectrum of the primaries\cite{Belavin:1984vu,Knizhnik:1984nr}.

As a different approach of classification, holomorphic modular bootstrap focuses on the modular properties of the characters $\operatorname{ch}$ (and torus correlation functions) by studying the modular differential equations they satisfy \cite{Mathur:1988na,Eguchi:1987qd,Bae:2020xzl,Das:2020wsi,Kaidi:2021ent,Das:2021uvd,Duan:2022kxr,Das:2022bxm,Mukhi:2022bte}. These ordinary differential equations are of the form
\begin{align}
	\left[D_q^{(n)} + \sum_{k = 0}^{n - 1}\phi_{2k}(\tau) D_q^{(k)}\right]\operatorname{ch} = 0 \ , \qquad q = e^{2\pi i \tau}\ ,
\end{align}
where $D^{(n)}_q$ denotes the Ramanujan-Serre derivatives, $\phi_{2k}$ are modular forms of weight $2n - 2k$ for $SL(2, \mathbb{Z})$ or suitable congruence subgroup if fermions are present. The equations themselves have relatively simple structure while bringing considerable constraints on the allowed characters, and therefore provides a useful organizing principal of RCFTs. The modular differential equation in modular bootstrap arises from the modular invariance of the RCFT, which states that the characters of the associated chiral algebra should form vector-valued modular form under the modular group. Automatically, the characters and their Ramanujan-Serre derivatives must satisfy an ordinary differential equation whose coefficients are modular forms.

In this paper, we explore a refinement of the modular bootstrap approach with \emph{flavored} modular differential equations, which we expect to apply to more general theories including non-rational ones. For RCFTs with flavor symmetry, different characters of different highest weight modules can have identical unflavoring limit, leading to degeneracy which may be lifted by flavor refinement. Moreover, highest weight modules of non-rational chiral algebras generally do not have finite dimensional weight space at a given conformal weight. In such cases, the unflavored limit of the characters does not exist, and therefore, they are simply invisible from looking at the unflavored modular differential equations. Therefore, flavor refinement improves the resolution and the range of validity of the holomorphic modular bootstrap method.

However, with flavor refinement and non-rationality in consideration, the derivation of modular differential equation based on Wronskian does not apply. Therefore we consider another circumstance where modular differential equations appear and serve as tool to classification of CFTs. In \cite{Gaberdiel:2008pr} it is shown that if a null state
\begin{equation}
	|\mathcal{N}\rangle = L_{-2}^s |0\rangle + \sum_{a, b} a_{ - h[a] - 1}|b\rangle 
\end{equation}
exists in the chiral algebra, it may give rise to a modular differential equation. In mathematical terminology, the null state implies that the stress tensor $T$ is nilpotent at the level of the $C_2$-algebra of the chiral algebra. This argument goes beyond rationality. For example, it is employed in the context of the 4d/2d correspondence where the relevant chiral algebra is in general non-rational. Every 4d $\mathcal{N} = 4$ superconformal field theory (SCFT) $\mathcal{T}$ contains a protected subsector of Schur operators that form a non-unitary and often non-rational\footnote{The associated chiral algebras $\chi(\mathcal{T})$ are expected to be quasi-lisse: a chiral algebra is quasi-lisse if its associated variety has finitely many symplectic leaves \cite{arakawa2012remark,Arakawa:2016hkg,Xie:2019vzr}.} associated chiral algebra $\chi(\mathcal{T})$ \cite{Beem:2013sza,Lemos:2014lua,Beem:2014rza,Song:2017oew,Xie:2019vzr,Xie:2019zlb,Xie:2019yds,Kiyoshige:2020uqz}. The Schur limit of the superconformal index of $\mathcal{T}$ is mapped under this correspondence to the vacuum character of $\chi(\mathcal{T})$, and the Higgs branch chiral ring is identified with the reduced Zhu's $C_2$-algebra \cite{Beem:2017ooy}. As a result, The non-Higgs-branch strong generators in $\chi(\mathcal{T})$ must be trivial at the level of the reduced Zhu's $C_2$-algebra. In particular, the stress tensor $T$ must be nilpotent in the $C_2$-algebra which is often implemented by a null state $\mathcal{N}_T$ in $\chi(\mathcal{T})$. Therefore by \cite{Gaberdiel:2008pr}, the unflavored Schur index of a 4d $\mathcal{N} = 2$ SCFT $\mathcal{T}$ is expected to satisfy some unflavored modular differential equation following from the nilpotency of the stress tensor \cite{Beem:2017ooy,Kaidi:2022sng,2023arXiv230409681L}, and this fact is used in the classification of a class of 4d $\mathcal{N} = 2$ SCFTs \cite{Kaidi:2022sng}. In the 4d/2d correspondence \cite{Beem:2013sza}, the presence of continuous flavor symmetry implies a non-trivial associated variety of $\chi(\mathcal{T})$ and therefore non-rationality of $\chi(\mathcal{T})$. Some of the non-vacuum characters of $\chi(\mathcal{T})$ are related to the Schur index of the $\mathcal{T}$ in the presence of certain surface defects \cite{Cordova:2017mhb,Beem:2017ooy,Nishinaka:2018zwq,Pan:2021ulr,Zheng:2022zkm}. Therefore, the intricate spectrum of the often non-rational $\chi(\mathcal{T})$ also encodes important information on the 4d physics.

The flavored modular differential equations are linear partial differential equations involving the derivatives $D_q^{(k)}$ (weight-two), $D_{b_i}$ (weight-one) and quasi-Jacobi coefficients $\phi_r(\mathfrak{b}_i, \tau)$. In \cite{Gaberdiel:2009vs}, flavored modular differential equations were applied to elliptic genera. In \cite{Peelaers,Zheng:2022zkm}, such flavored modular differential equations for some simple 4d $\mathcal{N} = 2$ Lagrangian SCFTs were studied, where additional solutions of those equations were found to have some 4d physical origins: the index of vortex surface defects, the residues of the integrand that computes the Schur index (which is related to Gukov-Witten type surface defects), and modular transformations of the Schur index (which is related to surface defects from singular background gauge fields coupled to the flavor symmetry).

To demonstrate some advantages of the flavor refinement, we will mainly focus for simplicity on a distinguished set of chiral algebras associated to the Deligne-Cvitanovi\'{c} series of exceptional Lie algebras \cite{Tuite:2014fha, Tuite:2008pt, Tuite:2006xi,Arakawa:2015jya,Arakawa:2016hkg,Beem:2013sza,Beem:2017ooy}
\begin{align}\label{Deligne-Cvitanovic}
	\mathfrak{g} = (\mathfrak{a}_0 \subset) \mathfrak{a}_1\subset \mathfrak{a}_2\subset \mathfrak{g}_2\subset \mathfrak{d}_4\subset \mathfrak{f}_4\subset \mathfrak{e}_{6} \subset \mathfrak{e}_7 \subset \mathfrak{e}_8\ .
\end{align}
This set of chiral algebras provides an ideal laboratory as it is very simple and at the same time spans a wide range of algebras, including the standard WZW models $\widehat{\mathfrak{g}}_{k = 1}$ with integral representations, Virasoro minimal model $M_{2,5}$, admissible Kac-Moody algebras such as $\widehat{\mathfrak{su}(2)}_{-4/3}$ and non-admissible ones $\widehat{\mathfrak{so}}(8)_{-2}, (\widehat{\mathfrak{e}}_6)_{-3}, (\widehat{\mathfrak{e}}_7)_{-4}, (\widehat{\mathfrak{e}}_8)_{-6}$.

To summarize, we will begin with a general Kac-Moody algebra $\widehat{\mathfrak{g}}_{k \ne - h^\vee}$ associated to a simple Lie algebra $\mathfrak{g}$. We then postulates a simplest null state $\mathcal{N}_T$ that implements the nilpotency of the stress tensor and satisfies $L_{n>0}|\mathcal{N}_T\rangle = 0$. From this, we explore the following aspects.
\begin{itemize}
	\item Using Zhu's recursion formula \cite{zhu1996modular,Mason:2008zzb,Peelaers,Pan:2021ulr} we derive the flavored modular differential equation that follow from the null state $|\mathcal{N}_T\rangle$, and the equation from the Sugawara stress tensor.
	\item We impose ``modularity'' on the two equations. The almost covariance \cite{Zheng:2022zkm} directly constrains the Lie algebra $\mathfrak{g}$ to the Deligne-Cvitanovi\'{c} series (\ref{Deligne-Cvitanovic}) and level $k$ to special values $k = 1, - h^\vee/6 - 1$.
	\item We discuss the relation between the ``modularity'' with the Joseph relations and nulls at weight-three. Roughly speaking, the $S$-transformation connects (the descendants of) nulls at different levels,
	\begin{equation}
		|\mathcal{N}_T\rangle \xrightarrow{S} |\mathcal{N}_T\rangle \oplus h^i_1|\mathcal{N}_T\rangle \oplus h^i_1h^j_1 |\mathcal{N}_T\rangle\ . 
	\end{equation}
	We argue that all highest weight flavored characters are determined by the two equations from $\mathcal{N}_T$ and $T = T_\text{Sug}$, and modularity \cite{Peelaers}.

	\item Finally, we explore the behavior of the equations under some shift $T_{\mathbf{n}^\vee}$ of the flavor fugacities $\mathfrak{b}_j$, which connects nulls at different levels,
	\begin{align}
		|\mathcal{N}_T\rangle \xrightarrow{T_{\mathbf{n}^\vee}} |\mathcal{N}_T\rangle \oplus h^i_1|\mathcal{N}_T\rangle \oplus h^i_1h^j_1 |\mathcal{N}_T\rangle \ .
	\end{align}
	Therefore we expect $T_{\mathbf{n}^\vee}$ to generate new solutions from the vacuum character.
\end{itemize}
From these discussions we see that the flavor refinement grants direct access to important internal structure and the representation theory of the chiral algebra, and is able to treat non-rational chiral algebras in much the same way as the rational ones.

This paper is organized as follows. In section \ref{section:prelim} we briefly recall some conventions and preliminaries on affine Lie algebra and chiral algebra. In section {\ref{section:FMLDEs}} we discuss the constraints on the chiral algebra and characters imposed by a weight-four null state, with the assumption of modularity. In section \ref{section:examples} we apply the discussions to a few examples.

\section{Preliminaries\label{section:prelim}}

In this paper we will focus on Kac-Moody Lie algebras, and let us begin by recalling some conventions and well-known results. Consider a simple Lie algebra $\mathfrak{g}$ of rank $r$. The generators of $\mathfrak{g}$ can be generically denoted as $J^A$ with the commutation relations $[J^A, J^B] = f^{AB}{_C} J^c$. The roots and simple roots of $\mathfrak{g}$ are denoted as $\alpha$ and $\alpha_i$, and the collection of roots $\Delta \coloneqq \{\alpha\}$. The Killing form $K(\cdot , \cdot)$ is defined by
\begin{align}
	K(X, Y) \coloneqq \frac{1}{2h^\vee} \operatorname{tr}_\text{adj} XY,  \qquad K^{AB} \coloneqq K(J^A, J^B)\ , \qquad \forall J^A, J^B, X, Y \in \mathfrak{g} \ .
\end{align}
The Killing form induces an inner product $(\cdot, \cdot)$ for the weights, such that for any long root $\alpha$, $|\alpha|^2 = (\alpha, \alpha) = 2$. The simple coroots are $\alpha_i^\vee \coloneqq 2\alpha_i/|\alpha_i|^2$ which are dual to the fundamental weights $\omega_i$. Similarly, the dual basis of the simple roots $\alpha_i$ are the fundamental coweights $\omega_i^\vee$, such that $(\alpha_i, \omega^\vee_j) = (\alpha_i^\vee, \omega_j) = \delta_{ij}$. The longest root and the Weyl vector are denoted by $\theta$ and $\rho$, and the dual Coxeter number $h^\vee \coloneqq (\theta, \rho) + 1$. The Cartan matrix of $\mathfrak{g}$ is given by $A_{ij} \coloneqq (\alpha_i, \alpha_j^\vee)$.

A Chevalley basis of generators of $\mathfrak{g}$ is denoted by $e^i, f^i, h^i$ with the (non-zero) commutation relations
\begin{align}
	[e^i, f^j] = \delta_{ij} h^j, \qquad
	[h^i, e^j] = A_{ji}e^j, \qquad
	[h^i, f^j] = - A_{ji}f^j \ , \qquad i = 1, \ldots, r \ .
\end{align}
In particular, $h^i$ span the Cartan subalgebra $\mathfrak{h} \subset \mathfrak{g}$. The remaining ladder operators $E^\alpha$ corresponding to roots $\alpha$ can be constructed by suitable nested commutators of $e$'s and $f$'s, and in particular, $e^i = E^{\alpha_i}$, $f^i = E^{- \alpha_i}$. In the end, all of the $h^i$ and $E^\alpha$ form a nice basis of $\mathfrak{g}$, with the commutation relations
\begin{align}
	[E^\alpha, E^{-\alpha}] = & \ \sum_{i = 1}^{r} f^{\alpha, -\alpha}{_i}h^i, \qquad
	[E^\alpha, E^\beta] = f^{\alpha \beta}{_\gamma} E^\gamma \text{ if } \gamma = \alpha + \beta \in \Delta \ , \\
	[h^i, E^\alpha] = & \ (\alpha, \alpha_i^\vee) E^\alpha \ .
\end{align}
In this Chevalley basis, the Killing form drastically simplifies, and the only non-zero components are
\begin{align}
	K^{ij} \coloneqq & \ K(h^i, h^j) = (\alpha_i^\vee, \alpha_j^\vee), \qquad \sum_{j = 1}^{r} K_{ij}K^{jk} = \delta_i^k, \\
	K^{\alpha, -\alpha} \coloneqq & \ K(E^{\alpha}, E^{-\alpha}), \qquad K_{\alpha, -\alpha} = (K^{\alpha, -\alpha})^{-1} \ .
\end{align}
In particular, for any two weights $\mu, \nu$
\begin{align}
	(\mu, \nu) = \sum_{i,j = 1}^{r} K^{ij}\mu_i \nu_i, \quad \text{where} \quad \mu = \sum_{i = 1}^{r} \mu_i\alpha_i^\vee, \quad \nu = \sum_{i = 1}^{r} \nu_i \alpha_i^\vee \ .
\end{align}
The structure constants are also related to $K$, 
\begin{align}
	f^{\alpha, -\alpha}{_i} = \frac{|\alpha_i|^2}{2} m^\alpha_i K^{\alpha, -\alpha} \ , \qquad
	f^{i \alpha}{_\alpha} = (\alpha, \alpha_i^\vee) \ , \qquad \alpha = \sum_{i = 1}^{r} m^\alpha_i \alpha_i \ .
\end{align}

The generators of an affine Kac-Moody algebra $\widehat{\mathfrak{g}}_k$ associated to $\mathfrak{g}$ are denoted as $J^A_n$ where $J^A = h^i, E^\alpha$, with the standard commutation relations
\begin{align}
	[J^A_m, J^B_n] = f^{AB}{_C} J^c_{m + n} + m k K^{AB} \delta_{m + n , 0} \ .
\end{align}
In terms of the vertex operators $J^A(z) = \sum_{n \in \mathbb{Z}} J^A_n z^{-n - 1}$, the above commutation relations are equivalent to the standard OPE
\begin{align}
	J^A(z) J^B(w) = \frac{f^{AB}{_C} J^C(w)}{z - w} + \frac{k K^{AB}}{(z - w)^2} + O(z - w) \ , \quad K^{AB} \coloneqq K(J^A, J^B) \ . \nonumber
\end{align}
For any non-critical level $k \ne - h^\vee$, one can define the Sugawara stress tensor $T_\text{Sug}(z)$ by the normal-ordered product,
\begin{align}
	T_\text{Sug}(z) = \frac{1}{2(k + h^\vee)} \sum_{A,B} K_{AB} (J^AJ^B)(z) = \sum_{n \in \mathbb{Z}} L_n z^{-n - 2} \ .
\end{align}
The modes $L_n$ satisfy the standard Virasoro commutation relations with a special central charge $c$,
\begin{align}
	[L_m, L_n] = (m - n)L_{m + n} + \frac{c}{12}m(m+1)(m - 1), \qquad
	c = \frac{k \dim \mathfrak{g}}{k + h^\vee} \ .
\end{align}
In other words, the affine Kac-Moody algebra $\widehat{\mathfrak{g}}_k$ contains a Virasoro subalgebra $V_c$ as long as $k$ is away from criticality. 

An affine primary $|\lambda\rangle$ is a state annihilated by $J_{n > 0}^A$ and $E^{\alpha > 0}_0$, with eigenvalues
\begin{align}
	h^i_{0} |\lambda\rangle = \lambda_i |\lambda \rangle, \qquad
	L_0 |\lambda\rangle = \frac{(\lambda, \lambda + 2 \rho)}{2(k + h^\vee)}|\lambda \rangle \ .
\end{align}
Starting from $|\lambda\rangle$ one can build a highest weight representation $M_\lambda$ of $\widehat{\mathfrak{g}}_k$ by acting with $J^A_{n < 0}$ and $E^{\alpha < 0}_0$, and in particular, the vacuum representation from $|0\rangle$. States in $M_\lambda$ besides $|\lambda\rangle$ are call affine descendants of $|\lambda\rangle$. If in a highest weight representation $M$ one finds a descendant state $|\mathcal{N}\rangle$ that also satisfies the affine primary condition, then $|\mathcal{N}\rangle$ is called an affine null state which signals the reducibility of $M_\lambda$ with a sub-representation generated by $|\mathcal{N}\rangle$. In general, any highest weight representation of $\widehat{\mathfrak{g}}_k$ also decomposes into irreducible representations of the Virasoro subalgebra $V_c$. In this sense, one can also define Virasoro null states that are annihilated by $L_{n > 0}$.

A null state $|\mathcal{N}\rangle$ in the vacuum representation of $\widehat{\mathfrak{g}}_k$ corresponds to a null field $\mathcal{N}(z)$ by the state-operator correspondence $a(z) \leftrightarrow |a\rangle$, where
\begin{align}
	a(z) = \sum_{n \in \mathbb{Z} - h_a} a_n z^{-n - h_a} \ , \qquad
	o(a) \coloneqq a_0 \ .
\end{align}
The torus one-point function of a null field (and its descendants) vanishes\footnote{Often we do not distinguish between nulls and descendants of null, as they play more or less the same role in the discussions.}. Such a one-point function can sometimes be pre-processed by the Zhu's recursion formula \cite{zhu1996modular,Mason:2008zzb,Beem:203X,Pan:2021ulr}. For a chiral algebra with a $U(1)$ affine current $J$, and any two operators $a(z), b(z)$ in the chiral algebra such that $J_0 |a \rangle = Q|a\rangle$,
\begin{align}
  \label{recursion2}
  \operatorname{str} o(a_{[- h_a]}b)x^{J_0}q^{L_0}
  = & \ \delta_{Q, 0}\operatorname{str}_M a_0 b_0 x^{J_0} q^{L_0} \\
  & \  + \sum_{n = 1}^{+\infty} E_n\left[ \begin{matrix}
    e^{2\pi i h_a} \\ x^Q
  \end{matrix} \right]
  \operatorname{str} o(a_{[- h_a+n]}b)x^{J_0}q^{L_0} \ ,
\end{align}
and when $n > 0$,
\begin{align}
	\operatorname{str} o(a_{[-h_a - n]} |b\rangle)q^{L_0} x^h
	= & \ (-1)^n \sum_{k = 1}^{+\infty} \begin{pmatrix}
	  k - 1\\n  
	\end{pmatrix}E_k \begin{bmatrix}
	  +1\\x^Q
	\end{bmatrix}\operatorname{tr}o(a_{[-h_a - n + k]}| b\rangle)q^{L_0}x^h \ . \nonumber
\end{align}
Here the square-mode $a_{[n]}$ follows from
\begin{align}
	a[z] \coloneqq e^{iz h_a} a(e^{iz} - 1) = \sum_{n \in \mathbb{Z} - h_a} a_{[n]} z^{-n - h_a} \ .
\end{align}

\section{Flavored modular differential equations\label{section:FMLDEs}}

\subsection{Nulls and FMDEs}

In \cite{Beem:2017ooy}, it is conjectured that the associated chiral algebra $\chi(\mathcal{T})$ of a 4d $\mathcal{N} = 2$ SCFT $\mathcal{T}$ encodes the Higgs branch of the latter as its associated variety, the presence of which signals non-rationality of the chiral algebra. As a result, the stress tensor $T$ of $\chi(\mathcal{T})$ should be nilpotent at the level of $C_2(\chi(\mathcal{T}))$,
\begin{align}
  L_{-2}^k |0\rangle = |\mathcal{N}_T\rangle + \varphi, \qquad \varphi \in C_2(\chi(\mathcal{T})) \ , \qquad k \in \mathbb{N}_{\ge 1}\ .
\end{align}
Here $\mathcal{N}_T$ is a null\footnote{State $ |\mathcal{N}_T\rangle$ can also be a descendant of some chiral null state, which is also removed to form a simple chiral algebra.} that bridges $T$ and $C_2(\chi(\mathcal{T})) $ is a vector space
\begin{align}
  C_2(\chi(\mathcal{T})) \coloneqq \operatorname{span}\{ a_{- h_a - 1}|b\rangle \ | \ a(z), b(z) \in \chi(\mathcal{T}) \} \ .
\end{align}
As a consequence of the nilpotency, it is conjectured that the Schur index should be a solution to a (unflavored) finite-order modular differential equation \cite{Beem:2017ooy,Arakawa:2016hkg}
\begin{align}
  \left[D_q^{(n)} + \sum_{k = 0}^{n - 1}\phi_r(\tau) D_q^{(r)} \right] \operatorname{ch} = 0 \ ,
\end{align}
where $\phi_r$ are $SL(2, \mathbb{Z})$ or $\Gamma^0(2)$ modular forms. This equation comes from computing the torus one-point function of $\mathcal{N}_T$. Such logic applies to all quasi-lisse chiral algebras, which include the more familiar rational chiral algebras as the simplest instances having zero dimensional associated varieties, and the null $\mathcal{N}_T$ is expected to play some central role in general.

Let us now focus on affine Kac-Moody algebras $\widehat{\mathfrak{g}}_k$. We assume the presence of a $\mathfrak{g}$-neutral weight-four affine or Virasoro null (or descendant of null) state $|\mathcal{N}_T\rangle$ in the vacuum module that is simultaneously Virasoro primary\footnote{If $|\mathcal{N}_T\rangle$ is also a Virasoro descendant, then $|\mathcal{N}_T\rangle$ is a usual Viraosro null state. This is the case in the $\mathfrak{a}_0$ entry of the non-unitary Deligne-Cvitanovi\'{c} series, where
\begin{equation}
  |\mathcal{N}_T\rangle = (L_{-2}^2 - \frac{5}{3}L_{-4})|0\rangle \ .
\end{equation}
}, which enforces the nilpotency of the stress tensor $T$. The general form of such a state is given by\footnote{Since $L_{-2}^2|0\rangle$ is neutral under $\mathfrak{g}$, the null $|\mathcal{N}_T\rangle$ is also expected to be neutral, hence all indices $ABC \ldots$ needs to contracted with invariant tensors. To bridge between $L_{-2}^2|0\rangle$ and $C_2(\widehat{\mathfrak{g}}_k)$, the remaining terms in $|\mathcal{N}_T\rangle$ should be of the form $a_{-h_a - 1}|b\rangle$, hence a term like $d_{ABCD} J^A_{-1}J^B_{-1}J^C_{-1}J^D_{-1}|0\rangle$ shall not appear.}
\begin{align}
  |\mathcal{N}_T\rangle = \left(L_{-2}^2 + \alpha L_{-4} + \beta K_{AB} J^A_{-3} J^B_{-1} + \gamma K_{AB}J^A_{-2} J^B_{-2} + \delta d_{ABC}J^A_{-2}J^B_{-1}J^C_{-1} \right) |0\rangle\ , \nonumber
\end{align}
where $d_{ABC}$ is the total symmetric cubic Casimir. In other words,
\begin{align}
  L_{-2}^2 |0\rangle = |\mathcal{N}_T\rangle - (\alpha L_{-4} + \beta K_{AB} J^A_{-3} J^B_{-1} + \gamma K_{AB}J^A_{-2} J^B_{-2} + \delta d_{ABC}J^A_{-2}J^B_{-1} J^C_{-1} ) |0\rangle \ , \nonumber
\end{align}
where the second term on the right belongs to vector space $C_2(\widehat{\mathfrak{g}}_k)$,
\begin{align}
  C_2(\widehat{\mathfrak{g}}_k) \coloneqq \operatorname{span}\{ a_{- h_a - 1}|b\rangle \ | \ a(z), b(z) \in \widehat{\mathfrak{g}}_k  \} \ .
\end{align}

For $|\mathcal{N}_T\rangle$ to be Virasoro primary, we have $L_{n > 0} |\mathcal{N}_T \rangle = 0$. Explicitly,
\begin{align}
  L_1 |\mathcal{N}_T \rangle = & \ \frac{3 + 5 \alpha + 3\beta (k + h^\vee) + 4\gamma(k + h^\vee)}{k + h^\vee}K_{ab} J^a_{-2}J^B_{-1}|0\rangle \nonumber\\
  & \ \qquad \qquad\qquad\qquad\qquad\qquad + 2\delta d_{ABC}J^A_{-1}J^B_{-1}J^C_{-1}|0\rangle = 0\ , \\
  L_2 |\mathcal{N}_T \rangle = & \ (8 + c + 6 \alpha + 6(h^\vee + k)\beta)L_{-2}|0\rangle = 0, \\
  L_{n \ge 3} |\mathcal{N}_T \rangle = & \ 0 \ .
\end{align}
They impose conditions on the coefficients $\alpha, \beta, \gamma, \delta$, 
\begin{align}
  \beta = - \frac{8 + c + 6 \alpha}{6(k + h^\vee)}, \qquad
  \gamma = \frac{2 + c - 4 \alpha}{8(k + h^\vee)} \ , \qquad \delta = 0 \ .
\end{align}
In particular,
\begin{align}
  \beta - 2 \gamma = - \frac{22 + 5c}{12 (k + h^\vee)} \ .
\end{align}

Besides $|\mathcal{N}_T\rangle$ at weight-four, there is another trivial null state $\mathcal{N}_\text{Sug}$ at weight-two that is expected to be present in any non-critical Kac-Moody algebra, corresponding to the statement that the stress tensor of the Kac-Moody algebra is given by the Sugawara stress tensor,
\begin{align}
  \mathcal{N}_\text{Sug}(z) = T(z) - T_\text{Sug}(z) = 0 \ .
\end{align}
This null field is also $\mathfrak{g}$-neutral like $\mathcal{N}_T$.

Let us now examine the flavored modular differential equations from the two null states. Explicitly, we consider the following equation of one-point function\footnote{Here the parity operator $(-1)^F$ is redundant for affine Kac-Moody algebra, but we keep it anyway. Also the $y$ factor the fugacity conjugate to the level $k$ \cite{Kac:1988qc}.},
\begin{align}
  0 = \langle \mathcal{N}(0)\rangle_M = y^k \operatorname{tr}_M \mathcal{N}(0) (-1)^F q^{L_0 - \frac{c}{24}} \prod_{i = 1}^{r} b_i^{h^i_0} \ ,
\end{align}
where $h^i$ are the Cartan generators in a Chevalley basis. In the absence of an insertion $\mathcal{N}(0)$, the trace gives the character of the module $M$,
\begin{align}
  \operatorname{ch}_M (b_i, q, y) = \operatorname{ch}_M (\mathfrak{b}_i, \tau, \mathfrak{y}) = y^k \operatorname{tr}_M (-1)^F q^{L_0 - \frac{c}{24}} \prod_{i = 1}^{r} b_i^{h^i_0} \ .
\end{align}
We often omit the subscript $M$ for brevity, and also write
\begin{align}
  q = e^{2\pi i \tau}, \qquad a_j = e^{2\pi i \mathfrak{a}_j}, \qquad b_j = e^{2\pi i \mathfrak{b}_j}, \qquad
  y = e^{2\pi i \mathfrak{y}} \ .
\end{align}
The corresponding derivatives are
\begin{align}
  q \partial_q = \frac{1}{2\pi i} \partial_\tau, \qquad
  D_{b_i} \coloneqq b_i \partial_{b_i} = \frac{1}{2\pi i} \partial_{\mathfrak{b}_i}, \qquad
  D_y \coloneqq y \partial_y = \frac{1}{2\pi i} \partial_\mathfrak{y} \ .
\end{align}

Applying Zhu's flavored recursion formula \cite{zhu1996modular, Gaberdiel:2009vs,bringmann2020zhu} to the one-point function of $\mathcal{N}_T$ leads to a partial differential equation,
\begin{align}\label{eqWeight4}
  0 = & \ D_q^{(2)}\operatorname{ch} +\left( \frac{c}{2} + 3kr (\beta - 2 \gamma) \right) E_4(\tau) \operatorname{ch}\\
  & \ + (\beta - 2 \gamma) \sum_{\alpha \in \Delta}\sum_{i = 1}^{r} K_{\alpha, -\alpha}f^{\alpha, -\alpha}{_i}E_3 \begin{bmatrix}
    1 \\ b^\alpha
  \end{bmatrix}D_{b_i}\operatorname{ch} + 3k (\beta - 2 \gamma) \sum_{\alpha \in \Delta} E_4 \begin{bmatrix}
    1 \\ b^\alpha
  \end{bmatrix}\operatorname{ch} \ . \nonumber
\end{align}
Here we denote
\begin{align}
  b^\alpha \coloneqq \prod_{j = 1}^r b_j^{\lambda_j^\alpha} \ ,\qquad
  \alpha = \sum_{i = 1}^{r}\lambda_i^\alpha \omega_i \in \Delta\ .
\end{align}
Similarly, the Sugawara null vector $\mathcal{N}_\text{Sug}$ leads to another equation,
\begin{align}\label{eqSugawara}
  0 = \Bigg( 2(k + h^\vee)D_q^{(1)}
  - K_{ij} D_{b_i}D_{b_j} 
  - \sum_{\alpha} K_{\alpha, - \alpha} & \ f^{\alpha, -\alpha}{_i} E_1 \begin{bmatrix}
      1 \\ b^\alpha
    \end{bmatrix}D_{b_i}\\
  & \ - k r E_2(\tau)
  - k \sum_{\alpha} E_2 \begin{bmatrix}
      1 \\ b^\alpha
    \end{bmatrix} \Bigg) \operatorname{ch} \ . \nonumber
\end{align}

\subsection{Modularity}

The two equations (\ref{eqWeight4}) and (\ref{eqSugawara}) are by no means the only equations that the characters should obey. In fact, we will argue that other equations (and nulls) are required to exist by modularity. As mentioned in the introduction, the Eisenstein series involved in the two flavored modular differential equations are not standard modular form of $SL(2, \mathbb{Z})$. Instead, they are quasi-Jacobi forms of $SL(2, \mathbb{Z})$. Let us consider the following $S$-transformation
\begin{align}
  \tau \to - \frac{1}{\tau}, \qquad
  \mathfrak{b}_i \to \frac{\mathfrak{b}_i}{\tau} \ ,\qquad
  \mathfrak{y} \to \mathfrak{y} - \frac{1}{k} \frac{1}{\tau} \sum_{i,j = 1}^r \mathcal{K}^{ij}\mathfrak{b}_i \mathfrak{b}_j \coloneqq \mathfrak{y} - \frac{1}{k} \frac{1}{\tau} \mathfrak{b}^2 \ .
\end{align}
Here $\mathcal{K}$ is a $r\times r$ symmetric matrix to be determined by modularity. It is easy to check that $S^2(\tau, \mathfrak{b}, \mathfrak{y}) = (\tau, - \mathfrak{b}, \mathfrak{y})$ is the charge conjugation, and $S^4 = \operatorname{id}$ on these variables.

The change of variables leads to mixture of derivatives,
\begin{align}
  q \partial_q \to & \ \tau^2 q \partial_q + \tau \sum_{i = 1}^{r} \mathfrak{b}_i D_{b_i} + \frac{1}{k} \sum_{i, j = 1}^{r} \mathcal{K}_{ij} \mathfrak{b}_i \mathfrak{b}_j D_y \nonumber \\
  = & \ \ \tau^2 q \partial_q + \tau \sum_{i = 1}^{r} \mathfrak{b}_i D_{b_i} + \sum_{i, j = 1}^{r} \mathcal{K}^{ij} \mathfrak{b}_i \mathfrak{b}_j \ ,\\
  D_{b_i} \to & \ \tau D_{b_i} + \frac{2}{k} \sum_{j = 1}^{r} \mathcal{K}^{ij} \mathfrak{b}_j D_y = \tau D_{b_i} + 2 \sum_{j = 1}^{r} \mathcal{K}^{ij} \mathfrak{b}_j \ .
\end{align}
Here we have anticipated that $D_y = k$ when acting on $y^k$. The Eisenstein series also transforms non-trivially, for example,
\begin{equation}
  E_3 \begin{bmatrix}
    1 \\ b^\alpha
  \end{bmatrix}
  \to
  \frac{i}{48 \pi^3} (\log b^\alpha)^3
    - \frac{\tau}{8\pi^2} E_1 \begin{bmatrix}
      1 \\ b^\alpha
    \end{bmatrix} (\log b^\alpha )^2
    + \frac{i \tau^2}{2\pi} E_2 \begin{bmatrix}
      1 \\ b^\alpha
    \end{bmatrix} \log(b^\alpha)
    + \tau^3 E_3 \begin{bmatrix}
    1 \\ b^\alpha
    \end{bmatrix} \nonumber \ ,
\end{equation}
where
\begin{align}
  \log b^\alpha = 2\pi i \sum_i \mathfrak{b}_i \lambda^\alpha_i = 2\pi i (\alpha, \mathfrak{b}), \qquad \mathfrak{b} \coloneqq \sum_{i} \mathfrak{b}_i \alpha^\vee_i \ .
\end{align}

Let us now impose modularity, by that we require the character to remain a solution to the $S$-transformed equations. We first explore its consequence with the $\mathcal{N}_\text{Sug}$ equation (\ref{eqSugawara}). After the $S$-transformation, we expect any module characters $\operatorname{ch}$ to stay the kernel of the following operator,
\begin{align}
  & \ 2(k + h^\vee)(\tau^2 D_q^{(1)} + \tau \mathfrak{b}_i D_{b_i} + \mathcal{K}^{ij} \mathfrak{b}_i \mathfrak{b}_j)
  - K_{ij}\Big(\tau D_{b_i} + 2 \mathcal{K}^{im}\mathfrak{b}_m \Big) \Big(\tau D_{b_j} + 2 \mathcal{K}^{jn}\mathfrak{b}_n \Big) \nonumber \\
  & \ - \sum_{\alpha} K_{\alpha, - \alpha}f^{\alpha, - \alpha}{_i} \left(\tau E_1 \begin{bmatrix}
    1 \\ b^\alpha  
  \end{bmatrix}
  + \lambda_j^\alpha \mathfrak{b}_j
  \right)(\tau D_{b_i} + 2 \mathcal{K}^{im}\mathfrak{b}_m)
  - kr \left(\tau^2 E_2 - \frac{\tau}{2\pi i}\right) \nonumber\\
  & \ - k \sum_{\alpha} \left(
  \tau^2 E_2 \begin{bmatrix}
    1 \\ b^\alpha  
  \end{bmatrix}
  - \tau (\lambda^\alpha_j \mathfrak{b}_j) E_1 \begin{bmatrix}
    1 \\ b^\alpha  
  \end{bmatrix}
  - \frac{1}{2} (\lambda_j^\alpha \mathfrak{b}_j)^2
  \right) \ .
\end{align}
Here repeated indices $i,j,m, n$ implies summation from $1$ to $r$.
Variables $\tau$ and $\mathfrak{b}$'s being arbitrary implies that the coefficients of $\tau^2$, $\tau \mathfrak{b}_i$ and $\mathfrak{b}_i \mathfrak{b}_j$ should separately annihilate the character. The $\tau^2$ part is obviously identical the original $\mathcal{N}_\text{Sug}$ equation (\ref{eqSugawara}). The $\mathfrak{b}_i \mathfrak{b}_j$ part is given by
\begin{align}
   2(k + h^\vee) \mathcal{K}^{ij} \mathfrak{b}_i \mathfrak{b}_j
  - & \ 4 K_{ij} \mathcal{K}^{im} \mathcal{K}^{jn} \mathfrak{b}_{m} \mathfrak{b}_n
   \nonumber \\
  & \ - 2\sum_{\alpha} K_{\alpha, -\alpha} f^{\alpha, - \alpha}{_i} \mathcal{K}^{im}\mathfrak{b}_m \lambda_n^\alpha \mathfrak{b}_n
  + \frac{k}{2} \sum_{\alpha}  \lambda_m^\alpha \mathfrak{b}_m  \lambda_n^\alpha \mathfrak{b}_n \ .
\end{align}
After a bit of rewriting using
\begin{equation}
  \sum_{\alpha}\lambda_m^\alpha \mathfrak{b}_m\lambda_n^\alpha \mathfrak{b}_n
  = \sum_{\alpha} (\alpha, \alpha_m^\vee)(\alpha, \alpha_n^\vee) \mathfrak{b}_m \mathfrak{b}_n = (\alpha_m^\vee, \alpha_n^\vee)\mathfrak{b}_m \mathfrak{b}_n
  = 2h^\vee K^{mn}\mathfrak{b}_m \mathfrak{b}_n \ ,
\end{equation}
and
\begin{equation}
  \sum_{\alpha} K_{\alpha, - \alpha}f^{\alpha, - \alpha}{_i} \mathcal{K}^{im} \mathfrak{b}_m \lambda_n^\alpha \mathfrak{b}_n
  = 2h^\vee \mathcal{K}^{im}\mathfrak{b}_n \ ,
\end{equation}
the $\mathfrak{b}_m \mathfrak{b}_n$ part simplifies to
\begin{align}
  &\ 2 (k + h^\vee) \mathcal{K}^{mn} \mathfrak{b}_m \mathfrak{b}_n
  - 4 K_{ij}\mathcal{K}^{im} \mathcal{K}^{jn}\mathfrak{b}_m \mathfrak{b}_n
  - 4 h^\vee \mathcal{K}^{mn}\mathfrak{b}_m \mathfrak{b}_n
  + k h^\vee K^{mn} \mathfrak{b}_m \mathfrak{b}_n \nonumber \\
  = & \ \mathfrak{b}^T K^{-1} (k \mathbf{1} - 2K \mathcal{K})(h^\vee + 2 K \mathcal{K}) \mathfrak{b} \ ,
\end{align}
where in the second line we adopted the matrix notation with $K \sim K_{mn}$, $K^{-1} \sim K^{mn}$ and $\mathfrak{b} \sim (\mathfrak{b}_1, \ldots, \mathfrak{b}_r)^T$. For this part to vanish, $\mathcal{K}$ is fixed to be
\begin{align}
  \mathcal{K}^{ij} = \frac{k}{2} K^{ij}, \qquad \text{or}, \qquad - \frac{h^\vee}{2} K^{ij} \ .
\end{align}
In other words, $\mathcal{K}^{ij} \sim \frac{\#}{2} K^{ij}$ for $\# = k$ or $- h^\vee$. Consequently, the $\tau \mathfrak{b}_j$ coefficient can be simplified to
\begin{align}
  2(k - \#)\bigg(D_{b_j} + \sum_{\alpha} \lambda_j^\alpha E_1 \begin{bmatrix}
    1 \\ b^\alpha  
  \end{bmatrix}\bigg) \ , \qquad \forall j = 1, \ldots, r\ . 
\end{align}
This finally fixes the $S$-transformation of $\mathfrak{y}$\footnote{The solution to the equation
\begin{align}
  \left(D_{b_j} + \sum_{\alpha}\lambda_j^\alpha E_1 \begin{bmatrix}
        1 \\ b^\alpha  
      \end{bmatrix}\right)\operatorname{ch} = 0
\end{align}
can be easily solved by $\operatorname{ch} = f(q) \prod_{\alpha} \vartheta_1(\alpha, \mathfrak{b})^{-1}$. However, this is not a solution to the original $\mathcal{N}_\text{Sug}$ equation.
}
\begin{align}
  \mathcal{K}^{ij} = \frac{k}{2} K^{ij} \ , \qquad
  \mathfrak{y} \xrightarrow{S} \mathfrak{y} - \frac{1}{\tau} \frac{1}{2} K^{ij} \mathfrak{b}_i \mathfrak{b}_j = \mathfrak{y} - \frac{1}{\tau} \frac{1}{2}(\mathfrak{b}, \mathfrak{b}), \quad \mathfrak{b} \coloneqq \sum_{i} \mathfrak{b}_i \alpha_i^\vee \ .
\end{align}

With this information, we move on to investigate the modularity of the $\mathcal{N}_T$-equation (\ref{eqWeight4}). By ``modularity'' we require the character to be annihilated by the following differential operator after $S$-transformation,
\begin{align}
  \xrightarrow{S} & \ \tau^4 D_q^{(2)} + 2\tau^3 \sum_{i} \mathfrak{b}_i D_{b_i}(D_q^{(1)} + E_2) + 2 \tau^2 \mathfrak{b}^2 (D_q^{(1)} + E_2)
  + \tau^2 \sum_{i, j} \mathfrak{b}_i \mathfrak{b}_j D_{b_i}D_{b_j}  \nonumber \\
  & \ + 2 \mathfrak{b}^2 \tau \sum_{i} \mathfrak{b}_i D_{b_i}
  + (\mathfrak{b}^2)^2 
  + \left(\frac{c}{2} + 3kr (\beta - 2 \gamma)\right) \tau^4 E_4(\tau) \nonumber\\
  & \ + (\beta - 2 \gamma)\sum_{\alpha \in \Delta} \sum_{i = 1}^{r} K_{\alpha, -\alpha}f^{\alpha, - \alpha}{_i} \Bigg[
  \frac{i}{48 \pi^3} (\log b^\alpha)^3
    - \frac{\tau}{8\pi^2} E_1 \begin{bmatrix}
      1 \\ b^\alpha
    \end{bmatrix} (\log b^\alpha )^2
    \nonumber\\
  & \ \qquad\qquad + \frac{i \tau^2}{2\pi} E_2 \begin{bmatrix}
      1 \\ b^\alpha
    \end{bmatrix} \log(b^\alpha)
    + \tau^3 E_3 \begin{bmatrix}
    1 \\ b^\alpha
  \end{bmatrix} \Bigg](\tau D_{b_i} + (\partial_{\mathfrak{b}_i} \mathfrak{b}^2))
   \\
  & \ + 3k (\beta - 2 \gamma)\sum_{\alpha \in \Delta} \Bigg[
  - \frac{1}{384 \pi^4} (\log b^\alpha)^4
    - \frac{i \tau}{48 \pi^3} (\log b^\alpha)^3 E_1 \begin{bmatrix}
      1 \\ b^\alpha
    \end{bmatrix} \nonumber \\
  & \ \qquad\qquad\qquad - \frac{\tau^2}{8 \pi^2} (\log b^\alpha)^2 E_2 \begin{bmatrix}
      1 \\ b^\alpha
    \end{bmatrix}
    + \frac{i\tau^3}{2\pi} (\log b^\alpha) E_3 \begin{bmatrix}
        1 \\ b^\alpha
      \end{bmatrix}
      + \tau^4 E_4 \begin{bmatrix}
        1 \\ b^\alpha
      \end{bmatrix}
    \Bigg] \nonumber \ .
\end{align}
Here $\mathfrak{b}^2 \coloneqq \mathcal{K}^{ij} \mathfrak{b}_i \mathfrak{b}_j = \frac{k}{2} (\mathfrak{b}, \mathfrak{b})$. The $\tau^4$ part of the transformed equation is identical to the original (\ref{eqWeight4}). Let us now focus on the $\tau^0$ part of the transformed equation, which reads
\begin{equation}
  (\mathfrak{b}^2)^2 + \frac{i (\beta - 2\gamma)}{48\pi^3}\sum_{\alpha \in \Delta}\sum_{i = 1}^{r}K_{\alpha, - \alpha}f^{\alpha, -\alpha}{_i} (\log b^\alpha)^3(\partial_{\mathfrak{b}_i} \mathfrak{b}^2) - \frac{ 3k (\beta - 2 \gamma)}{384\pi^4} \sum_{\alpha \in \Delta} (\log b^\alpha)^4 \ . \nonumber
\end{equation}
Some straightforward algebra simplifies the $\tau^0$-part to
\begin{align}
  \frac{k^2}{4}(\mathfrak{b}, \mathfrak{b})^2 + \frac{k}{24} (\beta - 2 \gamma) \sum_{\alpha} (\alpha, \mathfrak{b})^4 \ .
\end{align}
Note that the weight $\mathfrak{b}$ is arbitrary. It turns out for the $\tau^0$-part to vanish it imposes a very strong algebraic condition, that only the following simple Lie algebras are allowed:
\begin{align}
  \mathfrak{a}_1, \quad \mathfrak{a}_2, \quad \mathfrak{g}_2, \quad \mathfrak{d}_4, \quad \mathfrak{f}_4, \quad \mathfrak{e}_6, \quad \mathfrak{e}_7, \quad \mathfrak{e}_8 \ ,
\end{align}
which coincide with the well-known Deligne-Cvitanovi\'{c} exceptional series of simple Lie algebras. For this series, it is straightforward to check that
\begin{align}
  \sum_{\alpha} (\mathfrak{b}, \alpha)^4 = 6 \left(\frac{h^\vee}{6} + 1\right) (\mathfrak{b}, \mathfrak{b})^2\ , \qquad
  \forall \mathfrak{b} \ ,
\end{align}
and it forces
\begin{align}
  \frac{k}{4}\left[k + (\beta - 2\gamma)\left(\frac{h^\vee}{6} + 1\right)\right] = 0 \quad \Rightarrow \quad
  k = 0, \text{ or, } (\beta - 2 \gamma)\left(- \frac{h^\vee}{6} - 1\right) \ .
\end{align}
Recall that for $\mathcal{N}_T$ to be a Virasoro primary,
\begin{align}
  \beta - 2 \gamma = - \frac{22 + 5c}{12 (k + h^\vee)} \ ,
\end{align}

which implies a degenerate case with $c = - 22/5$ and $k = 0$: this is nothing but the $\mathfrak{a}_0$ entry of the Deligne-Cvitanovi\'{c} exceptional series corresponding to the Lee-Yang minimal model. If instead we assume $k \ne 0$ and the Sugawara central charge $c = k \dim \mathfrak{g}/(k + h^\vee)$, then the level $k$ is further constrained,
\begin{align}
  k = 1, \qquad - \frac{h^\vee}{6} - 1, \qquad - \frac{11}{6}h^\vee \ .
\end{align}
This recovers the level-one and non-unitary Deligne-Cvitanovi\'{c} exceptional series of affine Kac-Moody algebras. The Deligne-Cvitanovi\'{c} series have previously appeared in the context of vertex operator algebra and modular differential equations, see for example \cite{Tuite:2014fha, Tuite:2008pt, Tuite:2006xi,Arakawa:2016hkg,Beem:2017ooy}. The ADE entries of the non-unitary series all have 4d $\mathcal{N} = 2$ SCFT origin through the 4d/2d correspondence \cite{Beem:2013sza,Beem:2017ooy}. The last possibility $k = - \frac{11}{6} h^\vee$ can be excluded by additionally requiring the $|\mathcal{N}_T\rangle$ to partially satisfy the affine null condition\footnote{It is unclear to the authors how to justify such a requirement, or if there is other elegant argument to rule them out. In this paper we shall only focus on the former two solutions of $k$.}
\begin{align}
  J^A_{n = 2, 3} |\mathcal{N}_T\rangle = 0 \ .
\end{align}

Similarly, the $\tau^1$ part is given by
\begin{align}
  & \ 2 \mathfrak{b}^2 \sum_{i = 1}^{r} \mathfrak{b}_i D_{b_i}\\
  & \ + (\beta - 2 \gamma)\sum_{\alpha}\sum_{i = 1}^{r} \frac{|\alpha_i|^2}{2} m_i^\alpha\left(- \frac{(2\pi i \lambda^\alpha_j \mathfrak{b}_j)^2}{8\pi^2} E_1 \begin{bmatrix}
    1 \\ b^\alpha  
  \end{bmatrix} \mathcal{K}^{im} \mathfrak{b}_m
  + \frac{i (2\pi i \lambda_j^\alpha \mathfrak{b}_j)^3}{48\pi^3} D_{b_i}
  \right) \nonumber\\
  & \ - \frac{i}{48\pi^3} 3k (\beta - 2 \gamma) \sum_{\alpha}(2\pi i \lambda_j^\alpha \mathfrak{b}_j)^3 E_1 \begin{bmatrix}
    1 \\ b^\alpha  
  \end{bmatrix} \ ,
\end{align}
where all the $E_1$ terms cancel automatically, leaving
\begin{align}
  \left(k + \left(\frac{h^\vee}{6} + 1\right)(\beta - 2 \gamma)\right)\sum_{i} \mathfrak{b}_i D_{b_i} \ \Rightarrow \ k = - \left(\frac{h^\vee}{6} + 1\right)(\beta - 2 \gamma) \ .
\end{align}
But this is not a new constraint.

By modularity, the $\tau^3$ and $\tau^2$ part are additional flavored differential equations that the character should satisfy. Concretely, the $\tau^3$ part is 
\begin{align}\label{eq:e43}
  & \ 2  \sum_i \mathfrak{b}_i D_{b_i}(D_q^{(1)} + E_2)
  -  (\beta - 2 \gamma) \sum_{\alpha, i} \frac{|\alpha_i|^2}{2} m_i^\alpha  E_2 \begin{bmatrix}
    1 \\ b^\alpha  
  \end{bmatrix}(\alpha, \mathfrak{b}) D_{b_i} \nonumber \\
  & \ \qquad \qquad\qquad \qquad\qquad\qquad\qquad\qquad - 2 k  (\beta - 2\gamma) \sum_{\alpha} (\alpha, \mathfrak{b}) E_3 \begin{bmatrix}
    1 \\ b^\alpha  
  \end{bmatrix} \ ,
\end{align}
which implies four equations, and $\tau^2$ part is
\begin{align}\label{eq:e42}
  k & (\mathfrak{b}, \mathfrak{b}) (D_q^{(1)} + E_2)+ \mathfrak{b}_i \mathfrak{b}_j D_{b_i} D_{b_j}\\ 
  & \ + \frac{1}{2}(\beta - 2\gamma) \sum_{\alpha, i} \frac{|\alpha_i|^2}{2}m_i^\alpha (\alpha, \mathfrak{b})^2 E_1 \begin{bmatrix}
    1 \\ b^\alpha  
  \end{bmatrix}D_{b_i} 
  + \frac{1}{2}k (\beta - 2 \gamma) \sum_{\alpha} (\alpha, \mathfrak{b})^2 E_2 \begin{bmatrix}
    1 \\ b^\alpha  
  \end{bmatrix} \ , \nonumber
\end{align}
implying additional $\frac{r(r+1)}{2}$ equations. Under $S$-tranformation, (\ref{eq:e42}) transforms back to itself $\times \tau^2$, while the parts of lower weight vanish if $\mathcal{K}^{ij} = \frac{k}{2} K^{ij}$ and $k = - \left(\frac{h^\vee}{6} + 1\right)(\beta - 2\gamma)$. In a similar fashion, equation (\ref{eq:e43}) transforms into itself as the $\tau^3$ part, while the $\tau^2$ part reproduces (\ref{eq:e42}); parts of lower weight vanish if the above $\mathcal{K}$ and $k$ solution is applied. Therefore, the partial differential equation (\ref{eqSugawara}), (\ref{eqWeight4}), (\ref{eq:e43}) and (\ref{eq:e42}) form a closed system under the $S$-transformation: by modularity, they are expected to annihilate any module character $\operatorname{ch}$.

\subsection{Joseph ideal and highest weight characters}

All of the above equations are expected to come from additional null (descendant) states of the chiral algebra. Let us focus on the level $k = - \frac{h^\vee}{6} - 1$. At weight-two it is known that the affine generators $J^A$ satisfy the Joseph relations \cite{Arakawa:2015jya}\footnote{Here $\sim 0$ means the left hand side is an affine null or affine descendant. See all Theorem 4.2 of \cite{Arakawa:2015jya}.} 
\begin{align}\label{Joseph}
  (J^A J^B)|_{\mathfrak{R}} \sim 0 \ ,\qquad
  (J^A J^B)|_{\mathbf{1}} \sim K_{AB}(J^A J^B) \sim T \ .
\end{align}
where $\mathfrak{R}$ is a representation of $\mathfrak{g}$ in $\operatorname{symm}^2 \mathbf{adj} = \mathbf{1} \oplus \mathfrak{R} \oplus 2\mathbf{adj}$, and $\mathcal{I}_2 \coloneqq \mathfrak{R} \oplus \mathbf{1}$ is called the Joseph ideal. The representation $\mathfrak{R}$ for the Deligne-Cvitanovi\'{c} series are listed in Table \ref{table:R}.
\begin{table}
  \centering
  \begin{tabular}{c|c|c|c}
    $\mathfrak{g}$ & $\mathfrak{R}$ & number of charge-zero states in $\mathfrak{R}$ & $r(r+1)/2$\\
    \hline
    $\mathfrak{a}_1$ & $\mathbf{0}$ & 0 & $1$\\
    $\mathfrak{a}_2$ & $\mathbf{8}$ & 2 & $3$\\
    $\mathfrak{g}_2$ & $\mathbf{27}$ & 3 & $3$\\
    $\mathfrak{d}_4$ & $\mathbf{35}_\text{v} \oplus \mathbf{35}_\text{s} \oplus \mathbf{35}_\text{c}$ & 9  & $10$\\
    $\mathfrak{f}_4$ & $\mathbf{324}$ & 12  & $10$\\
    $\mathfrak{e}_6$ & $\mathbf{650}$ & 20  & $21$\\
    $\mathfrak{e}_7$ & $\mathbf{1539}$ & 27  & $28$\\
    $\mathfrak{e}_8$ & $\mathbf{3875}$ & 35  & $36$\\
  \end{tabular}
  \caption{List of $\mathfrak{R}$.\label{table:R}}

\end{table}

Within these Joseph relations, one can find null states uncharged under the Cartan $\mathfrak{h} \subset \mathfrak{g}$, and they shall give rise to non-trivial flavored modular differential equations of weight-two. From the table, we observe that for the ADE cases,
\begin{align}
  ADE: \qquad \# (\text{charge-zero states in } \mathfrak{R} ) + 1 = \frac{r(r+1)}{2} \ .
\end{align}
Therefore, we expect that in these cases the $\frac{r(r+1)}{2}$ weight-two equations in (\ref{eq:e42}) precisely correspond to those equations that arise from (\ref{Joseph}). This is indeed the case for $\mathfrak{a}_1, \mathfrak{a}_2, \mathfrak{d}_4$ by explicit construction of the null states. In other words, for the ADE cases in the Deligne-Cvitanovi\'{c} series with $k = - h^\vee/6 - 1$, the equation (\ref{eqWeight4}) alone is able to generate all the weight-two flavored modular differential equations including equation (\ref{eqSugawara}). For $\mathfrak{g}_2, \mathfrak{f}_4$, the equations (\ref{eq:e42}) are not enough to account for all the Joseph relations. For (\ref{eq:e42}) to contain (\ref{eqSugawara}), $\widehat{\mathfrak{g}}_k$ should satisfy $- \frac{kr}{2(k + h^\vee)} = 1$, which is indeed true for the ADE cases but not for $\mathfrak{g}_2, \mathfrak{f}_4$. It would be interesting to clarify the relation between the Joseph relations and the equations (\ref{eq:e42}) rigorously.

More concretely, we conjecture that the $r(r+1)/2$ equations (\ref{eq:e42}) and the $r$ equations (\ref{eq:e43}) arise from the zero-charge states $h^i_1 h^j_1 |\mathcal{N}_T\rangle, h^i_1 |\mathcal{N}_T\rangle$ at weight-two and three. Loosely speaking, the null states at different levels are connected by the $S$-transformation,
\begin{equation}
  |\mathcal{N}_T\rangle \xrightarrow{S} |\mathcal{N}_T\rangle \oplus h^i_1|\mathcal{N}_T\rangle \oplus h^i_1h^j_1 |\mathcal{N}_T\rangle\ . 
\end{equation}
For example,
\begin{align}
  J_1^A |\mathcal{N}_T\rangle = 2 J_{-1}^A L_{-2}|0\rangle + (\alpha + 1 + \beta k + 2 \gamma h^\vee)J_{-3}^A|0\rangle
  - (\beta - 2\gamma) f^A{_{BC}}J^B_{-2}J^C_{-1}|0\rangle \ . \nonumber
\end{align}
The torus one-point function of the above field with $J^A = h^i$ gives
\begin{align}
  \Big(2D_{b_i}D_q^{(1)} + 2 E_2(\tau)D_{b_i}
  - (\beta - 2\gamma)\sum_{\alpha,j} \frac{|\alpha_j|^2}{2} & \ m_j^\alpha (\alpha_i^\vee, \alpha) E_2 \begin{bmatrix}
    1 \\ b^\alpha  
  \end{bmatrix}D_{b_j}  \nonumber\\
  & \ - 2k (\beta - 2 \gamma)E_3 \sum_{\alpha} \begin{bmatrix}
    1 \\ b^\alpha  
  \end{bmatrix}  \Big) \operatorname{ch} \ , 
\end{align}
precisely reproducing (\ref{eq:e43}); note that $o(J_{[-3]}^A) = 0$. In particular, for the ADE entries with $k = - \frac{h^\vee}{6} - 1$, $J^A_1 J^B_1|\mathcal{N}_T\rangle|_\mathfrak{R}$ span the same subspace as $J^A_{-1}J^B_{-1}|0\rangle|_\mathfrak{R}$, while for other cases the latter is a larger subspace.

The system of equations (\ref{eqWeight4}), (\ref{eq:e43}), (\ref{eq:e42}) and (\ref{eqSugawara}) impose strong constraints on the allowed characters. Let us focus on the case with level $k = - \frac{h^\vee}{6} - 1$: we conjectured that this system is enough to fix the vacuum and all the highest weight module characters \footnote{Solving explicitly for $\mathfrak{a}_{1}, \mathfrak{a}_2, \mathfrak{g}_2$ and $\mathfrak{d}_4$ cases was carried out in \cite{Peelaers}. We thank Wolfger Peelaers for sharing his results.}. This can be achieved by assuming an anzatz
\begin{align}
  \operatorname{ch} = q^h \sum_{n = 0}^{+\infty}c_n(b_1, \ldots, b_r) q^n \ ,
\end{align}
and solving for $c_n(b_1, \ldots, b_r)$ order by order. To proceed, the weight-four equation (\ref{eqWeight4}) is actually the simplest one, because $D_{b_i}$ are all multiplied by $E_3 \big[ \substack{ + 1\\b} \big]$, which has the $q$-expansion
\begin{align}
  E_3 \begin{bmatrix}
    + 1 \\ b  
  \end{bmatrix} = - \frac{b^2 - 1}{2b} q + O(q^2) \ .
\end{align}
As a result, at the $n$-th order, $c_n(b_1, \ldots, b_r)$ always appears without derivatives, and $c_n$ is solved algebraically from $c_0, \ldots, c_{n-1}$: ultimately, all $c_{n \ge 1}$ are completely determined by $c_0$ alone. The weight-four equation (\ref{eqWeight4}) also implies only two solutions for $h$, $h = \frac{1}{12}(1+h^\vee)$ for the vacuum character and the other $h = \frac{1}{12}(1 - h^\vee)$ for the non-vacuum characters. For the vacuum character $\operatorname{ch}_0$, $c_0 = 1$ and one easily gets $\operatorname{ch}_0$ to arbitrary high $q$-order simply by applying equation (\ref{eqWeight4}). The task of solving the non-vacuum solutions $\operatorname{ch}$ reduces entirely to solving just one function $c_0(b_1,\ldots, b_r)$. The weight-three equations (\ref{eq:e43}) are all automatically satisfied at the zero-th order once the non-vacuum $h$ is plugged in. Concretely, using $E_2 \big[\substack{1 \\ b}\big] \sim E_2(\tau) \sim - \frac{1}{12} + O(q)$, $E_3 \big[\substack{1\\b}\big] \sim O(q)$, we see that the zero-th order of (\ref{eq:e43}) is proportional to
\begin{align}
  \left[2h - \frac{1}{6} + \frac{1}{6}h^\vee\right] D_{b_i} \operatorname{ch} \xrightarrow{h = \frac{1}{12}(1 - h^\vee)} 0, \qquad
  \operatorname{ch} = q^h c_0 (b_1, \ldots, b_r) \ .
\end{align}
Therefore, $c_0(b_1, \ldots, b_r)$ is constrained by the $r(r+1)/2$ equations of weight-two (and additionally (\ref{eqSugawara}) for the $\mathfrak{g}_2, \mathfrak{f}_4$ cases).

To solve $c_0$, one can further propose an anzatz for the function $c_0$,
\begin{align}
  c_0(b_1, \ldots b_r) = a_1^{n_1} \ldots a_r^{n_r} \sum_{\ell_i \ge 0} c_{0;\ell_1, \ldots, \ell_r} a_1^{\ell_1} \ldots a_r^{\ell_r}\ .
\end{align}
Here $a_i$ are fugacities that are conjugate to the simple roots,
\begin{align}
  a_i \coloneqq \prod_{j} b_j^{(\alpha_i, \alpha_j^\vee)} \ , \qquad
  D_{b_j} = \sum_{n = 1}^{r} \frac{|\alpha_n|^2}{2} K^{nj} D_{a_n} \ .
\end{align}
In the new variables $a_i$ and denoting $\mathbf{n} \coloneqq \sum_{j}n_j \alpha_j$, the equations (\ref{eq:e42}) at the leading order in $q$ and $a_i$ gives the algebraic equation
\begin{align}
  k\left[h - \frac{1}{12} - \frac{h^\vee}{12}(\beta - 2\gamma)\right] (\mathfrak{b}, \mathfrak{b})
  - \frac{1}{2}(\beta - 2\gamma) \sum_{\alpha > 0}(\mathbf{n}, \alpha)(\alpha, \mathfrak{b})^2
  + (\mathfrak{b}, \mathbf{n})^2 = 0\ , 
\end{align}
Additionally for $\mathfrak{g}_2, \mathfrak{f}_4$, equation (\ref{eqSugawara}) imposes at leading order
\begin{align}
  2(k + h^\vee) h - \sum_{i,j}(\mathbf{n},\mathbf{n})
  + \sum_{\alpha > 0} (\alpha, \mathbf{n}) + \frac{k}{12} \dim \mathfrak{g}  = 0 \ , \qquad \mathbf{n} \coloneqq \sum_{j}n_j \alpha_j \ .
\end{align}
For the non-vacuum value $h = \frac{1}{12} - \frac{h^\vee}{12}$, the above two sets of weight-two equations imply a finite number of solutions of $(n_1, \ldots, n_r)$. We list all the solutions in Table \ref{table:n}. We immediately recognize that these values of $n$ are precisely the (minus of) highest weights of the modules listed in the Table 1 of \cite{Arakawa:2015jya}, but written in the $\alpha_i$-basis.
\begin{table}
  \centering
  \begin{tabular}{c|c}
    $\mathfrak{g}$ & $\vec n$\\
    \hline
    $\mathfrak{a}_1$ & $(\frac{1}{3}), (\frac{2}{3})$\\
    $\mathfrak{a}_2$ & $(\frac{1}{2}, \frac{1}{2}), (1,\frac{1}{2}), (\frac{1}{2}, 1)$\\
    $\mathfrak{g}_2$ & $(2, \frac{4}{3}), (2, \frac{5}{3})$\\
    $\mathfrak{d}_4$ & $(1,2,1,1), (1,2,2,1), (1,2,1,2), (2,2,1,1)$\\
    $\mathfrak{f}_2$ & $(\frac{5}{2}, \frac{9}{2}, 6, 3), (3, \frac{9}{2}, 6, 3), (\frac{5}{2}, 5, 6, 3)$\\
    \hline
    $\mathfrak{e}_6$ & $(2, 4, 6, 4, 2, 3), (2, 4, 6, 5, 2, 3), (2, 4, 6, 5, 4, 3), $\\
    & $(2, 4, 6, 4, 2, 4), (2, 5, 6, 4, 2, 3), (4, 5, 6, 4, 2, 3)$\\
    \hline
    $\mathfrak{e}_7$ & $(4, 8, 12, 9, 6, 3, 6), (4, 8, 12, 9, 6, 3, 7), (4, 9, 12, 9, 6, 3, 6), (6, 9, 12, 9, 6, 3, 6),$\\
    & $ (4, 8, 12, 10, 6, 3, 6), (4, 8, 12, 10, 8, 3, 6), (4, 8, 12, 10, 8, 6, 6)$ \\
    \hline
    $\mathfrak{e}_8$ & $(10, 20, 30, 24, 18, 12, 6, 15), (10, 20, 30, 24, 18, 12, 6, 16),(10, 21, 30, 24, 18, 12, 6, 15), $\\
    & $(12, 21, 30, 24, 18, 12, 6, 15),(10, 20, 30, 25, 18, 12, 6, 15), (10, 20, 30, 25, 20, 12, 6, 15),$\\
    & $(10, 20, 30, 25, 20, 15, 6, 15), (10, 20, 30, 25, 20, 15, 10, 15)$
  \end{tabular}
  \caption{Solutions of $(n_1, \ldots, n_r)$ for $k = -h^\vee/6 - 1$.\label{table:n}}
\end{table}
Based on these observations, we made the following proposal.
\begin{itemize}
  \item For the $ADE$ Deligne-Cvitanovi\'{c} series with $k = - \frac{h^\vee}{6} - 1$, the weight-four equation (\ref{eqWeight4}) coming from the nilpotency of the stress tensor alone completely fix all the highest weight characters through modularity. These are chiral algebras have known 4d $\mathcal{N} = 2$ SCFT correspondence.

  \item For $\widehat{\mathfrak{g}}_2, \widehat{\mathfrak{f}}_4$ with $k = - \frac{h^\vee}{6} - 1$, the equation (\ref{eqWeight4}) and the weight-two equation (\ref{eqSugawara}) from the Sugawara construction together fix all the characters through modularity. These two chiral algebras do not have known associated $\mathcal{N} = 2$ SCFT.

  \item For the Deligne-Cvitanovi\'{c} series with $k = 1$, the equation (\ref{eqWeight4}) and additionally (\ref{eqSugawara}) from the Sugawara construction together also fix the characters of all the integral modules through modularity. Note that for $k = 1$ the equation (\ref{eqSugawara}) is outside of the $S$-orbit of (\ref{eqWeight4}).
\end{itemize}

The modularity of the flavored modular differential equations has another important implication: if $\operatorname{ch}(\mathfrak{b}, \tau, \mathfrak{y})$ is a module character, then its $S$-transformation $\operatorname{ch}(\mathfrak{b}/\tau, - 1/\tau, \mathfrak{y} - \frac{1}{\tau} \frac{1}{2} (\mathfrak{b}, \mathfrak{b}))$ is also a solution to all the above partial differential equations, and is likely some linear combination of the irreducible characters. For the representation theory of rational chiral algebras like the integral representations of the affine Kac-Moody algebras, or those with admissible levels, this conclusion is simply the well-known statement $\operatorname{ch}_i\xrightarrow{S} \sum_i S_{ij} \operatorname{ch}_j$ for a suitable modular $S$ matrix $S_{ij}$. 

However for the $\widehat{\mathfrak{d}}_4$, $\widehat{\mathfrak{e}}_{6,7,8}$ at non-admissible level $k = - \frac{h^\vee}{6} - 1$, the modular property of their characters is less understood: we claim that they still enjoy some similar modular property like the rest of the Deligne-Cvitanovi\'{c} exceptional series, with the subtlety that their characters are actually quasi-Jacobi forms, and logarithmic characters will be present in the $SL(2, \mathbb{Z})$-orbit of the vacuum character. This is known from the literature, \cite{Arakawa:2016hkg} for example, by studying the unflavored modular differential equation
\begin{align}
  (D_q^{(2)} - 5(h^\vee + 1) (h^\vee - 1)) \operatorname{ch}_0(q) = 0
\end{align}
satisfied by the unflavored vacuum character. The fact that it is second order indicates one logarithmic solution to the equation in the case of $\widehat{\mathfrak{d}}_4$, $\widehat{\mathfrak{e}}_{6,7,8}$ at level $k = - \frac{h^\vee}{6} - 1$. This also implies that the $SL(2, \mathbb{Z})$-orbit is relatively small and cannot generate all the solutions/characters of the Kac-Moody algebra: there are $r$ additional non-vacuum highest weight modules. To learn more about these other non-vacuum solutions/characters besides the logarithmic one, we now study the shift property of the flavored modular differential equations.

\subsection{Shift property and characters}

Besides almost $S$-covariance, the differential equations discussed previously further enjoy simple properties under shifts of the flavor fugacities $\mathfrak{b}_i$ and $\mathfrak{y}$. Let us consider shift
\begin{align}
  \mathfrak{b}_i \to \mathfrak{b}_i + n_i \tau, \qquad
  \mathfrak{y} \to \mathfrak{y} - \frac{1}{k} \sum_{i} \mathcal{K}^i (\mathfrak{b}_i + n_i \tau)
  - \frac{1}{k} \mathcal{K}_\tau \tau \ .
\end{align}
Here $\mathcal{K}$'s are constants to be determined by requiring almost covariance under the shift, and the numbers $n_i$ should satisfy
\begin{align}
  \sum_{i}n_i \lambda_i^\alpha = (\mathbf{n}^\vee, \alpha) \in \mathbb{Z}, \qquad \forall \alpha \in \Delta, \quad \mathbf{n}^\vee \coloneqq \sum_{i = 1}^{r} n_i \alpha_i^\vee \ .
\end{align}

Under the shift, the differential operators transform as
\begin{align}
  q \partial _q \to q \partial_q - \sum_{i} n_i D_{b_i} + \frac{1}{k} \mathcal{K}_\tau D_y \ ,
  \quad
  D_{b_i} = D_{b_i} + \mathcal{K}^i \ .
\end{align}
The $\mathcal{N}_\text{Sug}$ equation transforms into
\begin{align}
  \to & \ 2(k + h^\vee) \left(D_q^{(1)} - \sum_{i = 1}^{r} n_i D_{b_i} + \mathcal{K}_\tau\right)
  - \mathcal{K}_{ij} (D_{b_i} + \mathcal{K}_i) (D_{b_j} + \mathcal{K}_j) \nonumber \\
  & \ - \sum_{\alpha, i} \frac{|\alpha_i|^2}{2} m^\alpha_i \left(
    E_1 \begin{bmatrix}
      1 \\ b^\alpha
    \end{bmatrix}
    - n_j \lambda_j^\alpha
  \right)(D_{b_i} + \mathcal{K}_i)\\
  & \ - k \sum_{\alpha} \left(
  E_2 \begin{bmatrix}
    1 \\ b^\alpha  
  \end{bmatrix}
  + n_j \lambda_j^\alpha E_1 \begin{bmatrix}
    1 \\ b^\alpha  
  \end{bmatrix}
  - \frac{1}{2}(n_j \lambda^\alpha_j)^2
  \right) \ . \nonumber
\end{align}
The weight-one part reads
\begin{align}
  & \ -2(k + h^\vee)n_i D_{b_i} - 2 K_{ij} \mathcal{K}^i D_{b_j} + \sum_{\alpha, i}\frac{|\alpha_i|^2}{2} m^\alpha_i n_j\lambda^\alpha_j D_{b_i} \nonumber\\
  & \ - \sum_{\alpha,i} \frac{|\alpha_i|^2}{2} m_i^\alpha \mathcal{K}^i E_1 \begin{bmatrix}
    1 \\ b^\alpha  
  \end{bmatrix} 
  - k \sum_{\alpha} n_j\lambda_j^\alpha E_1 \begin{bmatrix}
    1 \\ b^\alpha  
  \end{bmatrix} \\
  = & \ -2(k n_j + K_{ij} \mathcal{K}^i)D_{b_j}
  - \sum_{\alpha, i} \frac{|\alpha_i|^2}{2} m_i^\alpha \left(\mathcal{K}^i + k n_j K^{ij}\right) E_1 \begin{bmatrix}
    1 \\ b^\alpha  
  \end{bmatrix} \ . \nonumber
\end{align}
Again, this equation can only be satisfied if
\begin{align}
  \mathcal{K}^i + k n_j K^{ij} = 0 \ .
\end{align}
Similarly, the weight-zero part of the transformed equation is
\begin{align}
  2(k + h^\vee) \mathcal{K}_\tau + (k + h^\vee) n_i \mathcal{K}^i \quad 
  \Rightarrow \quad
  \mathcal{K}_\tau = - \frac{1}{2} \sum_{i = 1}n_i \mathcal{K}^i
  = + \frac{k}{2} \sum_{i}n_i K^{ij} n_j \ .
\end{align}
To summarize, the shift transformation is fixed to be
\begin{align}
  \mathfrak{b}_i \to \mathfrak{b}_i + n_i \tau, \qquad
  \mathfrak{y} \to \mathfrak{y} + \sum_{i} K^{ij}n_j \mathfrak{b}_i
  + \frac{1}{2} K^{ij}n_i n_j \tau\ ,
\end{align}
and under this shift, the equation (\ref{eqSugawara}) transforms back to itself.

The equation (\ref{eqWeight4}) similarly transforms into itself and some weight-zero, -one, -two and -three parts. In particular, the weight-zero and weight-one parts vanish identically when the above solution of $\mathcal{K}_\tau$, $\mathcal{K}_i$ is applied. On the other hand, the $\tau^2$ and $\tau^3$ part are precisely the equation (\ref{eq:e42}) and (\ref{eq:e43}) that appear in the $S$-transformed $\mathcal{N}_T$, and they were all required to be satisfied by the character from the modularity assumption.

Similarly, it is also straightforward to show that the same shift property is true for equation (\ref{eq:e43}) and (\ref{eq:e42}) that appear in the $S$-transformed (\ref{eqWeight4}).

The direct conclusion of the above, combined with the modularity assumption, is that the shifted characters
\begin{align}
  \operatorname{ch}(b_i, q, y) \to \operatorname{ch}(b_i q^{n_i}, q, y \prod_{i}b_i^{K^{ij}n_j} q^{+ \frac{1}{2}K^{ij}n_i n_j} )
\end{align}
remain solutions to all the equations (\ref{eqWeight4}), (\ref{eqSugawara}), (\ref{eq:e43}) and (\ref{eq:e42}). In terms of $\mathfrak{b} \coloneqq \sum_i \mathfrak{b}_i \alpha_i^\vee$, $\mathbf{n}^\vee = \sum_{i} n_i \alpha^\vee_i$, the shift can also be written as
\begin{align}\label{T-operator}
  \operatorname{ch}(\mathfrak{b}, \tau, \mathfrak{y})
  \to T_{\mathbf{n}^\vee}\operatorname{ch}(\mathfrak{b}, \tau, \mathfrak{y})
  \coloneqq 
  \operatorname{ch}(\mathfrak{b} + \mathbf{n}^\vee \tau, \tau, \mathfrak{y} + (\mathfrak{b}, \mathbf{n}^\vee) + \frac{1}{2}(\mathbf{n}^\vee, \mathbf{n}^\vee)\tau) \ .
\end{align}
This result is well-known for integrable and admissible characters, and we are proposing that it generalizes to the non-admissible cases in the Deligne-Cvitanovi\'{c} series.

\section{Examples\label{section:examples}}

\subsection{$\widehat{\mathfrak{su}}(2)_{-4/3}$}

Let us elaborate in this simplest example $\widehat{\mathfrak{su}}(2)_{-4/3}$. It is the associated chiral algebra of the Argyres-Douglas theory $(A_1, D_3)$ in four dimensions in the infinite series $(A_1, D_{2n + 1})$ \cite{Argyres:1995jj,Xie:2012hs,Song:2017oew,Creutzig:2017qyf}. We take the basis of $\mathfrak{su}(2)$ to be
\begin{align}
	J^- = \begin{pmatrix}
  	0 & 0 \\ 
  	1 & 0  
	\end{pmatrix}, \qquad
	J^3 = \frac{1}{2} \begin{pmatrix}
  	1 \\ & -1
	\end{pmatrix}, \qquad
	J^+ = \begin{pmatrix}
  	0 & 1 \\
  	0 & 0  
	\end{pmatrix} \ .
\end{align}
The adjoint representation $\mathfrak{su}(2)$ is $\mathbf{3}$, and
\begin{align}
	\mathbf{3} \otimes \mathbf{3} = \mathbf{1} \oplus \mathbf{3} \oplus \mathbf{5} = \mathbf{1} \oplus \wedge^2 \mathbf{3} \oplus 2\mathbf{adj}, \qquad \mathfrak{R} = \mathbf{0} \ .
\end{align}
Therefore, apart from the equation from the Sugawara construction $\mathcal{N}_\text{Sug} = T - T_\text{Sug}$, there is no additional null state from the Joseph relations. At weight-three there is one affine null state given by
\begin{equation}
	|\mathcal{N}_3^+\rangle 
	= \left(- \frac{13}{9} J_{-3}^+
	+ \frac{11}{3} J^3_{-2}J^+_{-1}
	- \frac{2}{3}J^+_{-2}J^3_{-1}
	+ J^-_{-1}J^+_{-1} J^+_{-1}
	+ J^+_{-1} J^3_{-1}J^3_{-1}
	\right)|0\rangle .
\end{equation}
Under the finite $\mathfrak{su}(2)$ action, it generates a representation $\mathbf{3}$ with two other null descendants at the same level,
\begin{align}
	|\mathcal{N}_3^3 \rangle
	= & \ \big(- \frac{22}{9}J^3_{-3}
	+ J^3_{-2} J^3_{-1}
	- \frac{4}{3} J^-_{-2}J^+_{-1}
	+ \frac{4}{3} J^+_{-2}J^-_{-1}
	+ J^3_{-1}J^3_{-1}J^3_{-1}
	+ J^-_{-1}J^+_{-1}J^3_{-1}
	\big)|0\rangle \nonumber \\
	= & \ \left(L_{-2}h_{-1}
	- e_{-1}f_{-2} + e_{-2} f_{-1} - \frac{2}{3}h_{-3}
	\right) |0\rangle \ , \nonumber \\
	|\mathcal{N}_3^-\rangle 
	= & \ \left(+ \frac{5}{9} J^-
	+ \frac{1}{3} J^3_{-2}J^-_{-1}
	+ \frac{2}{3}J^-_{-2}J^3_{-1}
	+ J^-_{-1}J^-_{-1} J^+_{-1}
	+ J^-_{-1} J^3_{-1}J^3_{-1}
	\right)|0\rangle \ . \nonumber
\end{align}
It is easy to verify that
\begin{align}
	 J^A_{n > 0}|\mathcal{N}_3^A\rangle = 0\ , \qquad
	\langle \mathcal{N}_3^A | \mathcal{N}_3^B\rangle = 0 \ , \qquad
	|\mathcal{N}_3^A \rangle = J^A_1 |\mathcal{N}_T\rangle \ ,
\end{align}
and the zero-charge null descendant $|\mathcal{N}_3^3 \rangle$ corresponds to a weight-three flavored modular differential equation. At weight-four, there is $|\mathcal{N}_T\rangle$ that enforces the nilpotency of $T$, giving rise to the weight-four modular differential equation. Note also that $|\mathcal{N}_T\rangle$ is a null descendant of $|\mathcal{N}_3^+\rangle$,
\begin{align}
	|\mathcal{N}_T\rangle = \frac{9}{8} \Big(2J^3_{-1} |\mathcal{N}_3^3\rangle
	+ J^-_{-1} |\mathcal{N}_3^+\rangle
	+ J^+_{-1} |\mathcal{N}_3^-\rangle
	\Big) \ .
\end{align}
All of $|\mathcal{N}_3^A\rangle$ and $|\mathcal{N}_T\rangle$ are annihilated by $L_{n > 0}$, and $J^3_1 J^3_1 |\mathcal{N}_T\rangle = 0$ is consistent with $\mathfrak{R} = \mathbf{0}$.

Concretely, the equations from the above nulls are
\begin{equation}\label{su2Weight2}
	\left[
	D_q^{(1)} - \frac{3}{8} D_{b_1}^2
	- \frac{3}{2}E_1 \begin{bmatrix}
	  	1 \\ b_1^2  
		\end{bmatrix}D_{b_1}
		+ 2E_2 \begin{bmatrix}
	  	1 \\ b_1^2  
		\end{bmatrix}
		+ E_2 (\tau)
	\right] \operatorname{ch} = 0 \ ,
\end{equation}
\begin{equation}\label{su2Weight3}
	\left[D_{b_1}D_q^{(1)} - \left( 2 E_2 \begin{bmatrix}
		  	1 \\ b_1^2  
			\end{bmatrix} - E_2(\tau) \right)D_{b_1}
			+ \frac{16}{3} E_3 \begin{bmatrix}
		  	1 \\ b_1^2  
			\end{bmatrix}\right]\operatorname{ch} = 0 \ ,
\end{equation}
\begin{equation}\label{su2Weight4}
	\left[D_q^{(2)} + 2 E_3 \begin{bmatrix}
		  	1 \\ b_1^2  
			\end{bmatrix} D_{b_1} - 8 E_4 \begin{bmatrix}
		  	1 \\ b_1^2  
			\end{bmatrix} - 7 E_4(\tau)\right]\operatorname{ch} = 0 \ .
\end{equation}
These three equations form a closed system under $S$-transformation, and in particular both (\ref{su2Weight2}) and (\ref{su2Weight3}) sit in the $S$-orbit of (\ref{su2Weight4}),
\begin{align}
	S\text{(weight-4)} = \tau^4\text{(weight-4)} + 2 \tau^3\mathfrak{b}_1\text{(weight-3)} + \frac{8}{3}\tau^2 \mathfrak{b}_1^2\text{(Sugawara)} \ .
\end{align}

Although there are more equations at these conformal weights that comes from the descendants of the lower-weight nulls \footnote{For example, at weight-three, there is another equation
\begin{align}
	\Bigg[D_{b_1}^3 + \frac{32}{3} D_q^{(1)} E_1 \begin{bmatrix}
  	1 \\ b_1^2  
	\end{bmatrix}
	- 8 \bigg(3 E_1 \begin{bmatrix}
  	1 \\ b_1^2  
	\end{bmatrix}^2 & \ + \frac{10}{3}E_2 \begin{bmatrix}
  	1 \\ b_1^2  
	\end{bmatrix} + E_2 (\tau)\bigg) D_{b_1} \nonumber\\
	& \ + 32 E_1 \begin{bmatrix}
  	 1 \\ b_1^2  
	\end{bmatrix}E_2 \begin{bmatrix}
  	1 \\ b_1^2  
	\end{bmatrix}
	+ \frac{416}{9}E_3 \begin{bmatrix}
  	1 \\ b_1^2  
	\end{bmatrix}\Bigg] \operatorname{ch} = 0 \ .
\end{align}
The $S$-orbit of this equation also contains the equation (\ref{su2Weight2}) from the Sugawara construction, and its weight-one and zero part also fixes $\mathcal{K}^{ij} = \frac{k}{2}K^{ij}$.
}, the above three equations (or equivalently, (\ref{su2Weight4}) alone with modularity) are enough to fix all three highest weight characters \cite{Peelaers}. To proceed, one can make an anzatz for the solution
\begin{align}
	\operatorname{ch} = q^h \sum_{n = 0}^{+\infty} c_n(b_1) q^n \ ,
\end{align}
and solve the coefficient function $c_n(b)$ order by order. For example, at leading order, equation (\ref{su2Weight4}) fixes $h = - \frac{1}{12}$ or $h = \frac{1}{4}$. Then equation (\ref{su2Weight3}) and (\ref{su2Weight2}) respectively instructs that if $h \ne - \frac{1}{12}$ then $c_0(b_1)$ must be constant, and if $h = - \frac{1}{12}$, $c_0(b_1) $ must satisfy
\begin{align}
	9 b_1^2 c''_0(b_1) + 8c_0(b_1) + \frac{9b_1 (1 + b_1^2) c_0'(b_1)}{b_1^2 - 1} = 0 \ .
\end{align}
with solutions given by
\begin{align}
	c_0(b_1) = c_1 \frac{b_1^{2/3}}{1 - b_1^2} + c_2 \frac{b_1^{4/3}}{1 - b_1^2} \ , \qquad c_1, c_2 \text{ are constants}.
\end{align}
From this we can read off three possibilities,
\begin{align}
	\operatorname{ch} = q^{\frac{1}{4}}(1 + \cdots), \quad
	q^{- \frac{1}{12}}\left(\frac{b_1^{2/3}}{1 - b_1^2}
	+ \cdots
	\right) , \quad
	q^{- \frac{1}{12}}\left(\frac{b_1^{4/3}}{1 - b_1^2}
		+ \cdots
		\right) .
\end{align}
Once the leading term is fixed to be either one of the above, higher order terms are completely determined by (\ref{su2Weight4}) alone. Obviously, they are precisely the three admissible characters with highest weights
\begin{align}
	- \frac{4}{3} \widehat{\omega}_0, \qquad
	- \frac{4}{3} \widehat{\omega}_1, \qquad
	- \frac{2}{3} \widehat{\omega}_0 - \frac{2}{3} \widehat{\omega}_1 \ .
\end{align}
These highest weight characters respectively can be written in well-known closed forms,
\begin{align}
	\operatorname{ch}_0 = & \ y^{-4/3} \frac{\vartheta_1(2 \mathfrak{b}| 3\tau)}{\vartheta_1(2 \mathfrak{b} | \tau)} \ , \\
	\operatorname{ch}_1 = & \ y^{-4/3} b^{-2/3} q^{\frac{1}{6}} \frac{\vartheta_1(2 \mathfrak{b} - \tau| 3\tau)}{\vartheta_1(2 \mathfrak{b} | \tau)} \ , \\
	\operatorname{ch}_2 = & \ y^{-4/3} b^{- 4/3} q^{\frac{2}{3}} \frac{\vartheta_1(2 \mathfrak{b} - 2\tau| 3\tau)}{\vartheta_1(2 \mathfrak{b} | \tau)} \ .
\end{align}
Obviously,
\begin{equation}
	\operatorname{ch}_n(\mathfrak{b}, \tau, \mathfrak{y}) = (-1)^n e^{2\pi i (\mathfrak{b}, \mathbf{n}^\vee)} e^{\pi i k (\mathbf{n}^\vee, \mathbf{n}^\vee)\tau} \operatorname{ch}(\mathfrak{b} + \mathbf{n} \tau, \tau, \mathfrak{y}) \ , \qquad
	\mathbf{n}^\vee = n \alpha_1^\vee \ .
\end{equation}

\subsection{$\widehat{\mathfrak{su}}(2)_{k = 1}$}

At integral level $k = 1$, the algebra contains a well-known affine null state at level-two $\mathcal{N}_2^\text{hw} \coloneqq J^+_{-1}J^+_{-1}|0\rangle$ which generates a subspace $\mathbf{5}$ with four other null descendants at the same level. The zero charge state $\mathcal{N}_2^0 \coloneqq (J^3_{-2} - 2 J^3_{-1}J^3_{-1} + J^-_{-1} J^+_{-1})|0\rangle$ leads to an additional flavored modular differential equation besides (\ref{eqSugawara}).

The level-four state
\begin{align}
	|\mathcal{N}_T\rangle \coloneqq
	\Big(L_{-2}^2 - \frac{1}{2} K_{AB}J^A_{-3}J^B_{-1} + \frac{1}{8}K_{AB}J^A_{-2}J^B_{-2} \Big)|0\rangle
\end{align}
enforces the nilpotency of $T$ and is also an affine descendant of $\mathcal{N}_2^\text{hw}$. The level-two states $J^A_1 J^B_1| \mathcal{N}_T\rangle$ also span the subspace $\mathbf{5}$ led by $|\mathcal{N}_2^\text{hw}\rangle$, and the states $J^3_1|\mathcal{N}_T\rangle $, $J^3_1 J^3_1|\mathcal{N}_T\rangle = - \frac{6}{7}\mathcal{N}^0_2$ give rise to the weight-three and weight-two equation (\ref{eq:e43}), (\ref{eq:e42}). Concretely, the equation from the Sugawara condition is
\begin{align}
	\bigg(D_q^{(1)} - \frac{1}{12}D_{b_1}^2
	- \frac{1}{3} E_1 \begin{bmatrix}
  	1 \\ b_1^2  
	\end{bmatrix} D_{b_1}
	- \frac{1}{6} E_2 - \frac{1}{3}E_2 \begin{bmatrix}
  	1 \\ b_1^2  
	\end{bmatrix}\bigg)\operatorname{ch} = 0 \ .
\end{align}
Equations (\ref{eq:e42}), (\ref{eq:e43}) and (\ref{eqWeight4}) are given explicitly by
\begin{align}
	0 = & \ \bigg(
	D_q^{(1)} + \frac{1}{2}D_{b_1}^2 - \frac{3}{2} E_1 \begin{bmatrix}
  	1 \\ b_1^2  
	\end{bmatrix}D_{b_1} + E_2(\tau) - \frac{3}{2}E_2 \begin{bmatrix}
  	1 \\ b_1^2  
	\end{bmatrix}
	\bigg) \operatorname{ch} \ , \\
	0 = & \ \bigg(
	D_{b_1}D_q^{(1)} + \Big(E_2 + \frac{3}{2} E_2 \begin{bmatrix}
	  	1 \\ b_1^2  
		\end{bmatrix} \Big)D_{b_1} + 3 E_3 \begin{bmatrix}
	  	1 \\ b_1^2  
		\end{bmatrix}
	\bigg)\operatorname{ch} \ ,\\
	0 = & \ \bigg(D_q^{(2)} - \frac{3}{2} E_3 \begin{bmatrix}
  	1 \\ b_1^2  
	\end{bmatrix}D_{b_1} - \frac{7}{4} E_4 - \frac{9}{2} E_4 \begin{bmatrix}
  	1 \\ b_1^2  
	\end{bmatrix}\bigg) \operatorname{ch} \ .
\end{align}
These three equations form a closed subset under the $S$-transformation, while the Sugawara equation is outside of the $S$-orbit. The four equations above completely fix the highest weight characters starting from an anzatz $\operatorname{ch} = q^h a_0(b_1)$. The weight-four equation fixes $h = -1/24, 5/24$. The two weight-two equations together impose constraints
\begin{align}
	a'_0(b_1) = & \ 0\ , \ & \text{ if } \ h = & \ - \frac{1}{24}, \\
	a_0'(b_1) = & \ \frac{b_1^2 - 1}{b_1(1 + b_1^2)}a_0(b_1) \ , \ & \text{ if } \ h = & \ + \frac{5}{24} \ .
\end{align}
The former is simply the vacuum solution, while the second case gives the integral character of highest weight $\widehat{\omega}_1$,
\begin{align}
	a_0(b_1) = b_1 + \frac{1}{b_1} = \chi_{\mathbf{2}}(\mathfrak{su}(2)) \ .
\end{align}

\subsection{\texorpdfstring{$\widehat{\mathfrak{su}}(3)_{-3/2}$}{}}

The level $k = - 3/2$ is boundary admissible with respect to $\mathfrak{g} = \mathfrak{su}(3)$. At level-two, besides the Sugawara condition $T - T_\text{Sug}$, there is one affine null state given by
\begin{align}
	|\mathcal{N}^\text{hw}_2\rangle = \Big(3 E^{\alpha_1 + \alpha_2}_{-2} + 6 E^{\alpha_1}_{-1}E^{\alpha_2}_{-1} + 2 E^{\alpha_1 + \alpha_2}_{-1}(h^1_{-1} - h^2_{-1}) \Big)|0\rangle \ .
\end{align}
It generates an $\mathfrak{R} = \mathbf{8}$ representation with seven affine descendants at the same level that implement the Joseph relations $J^AJ^B|_\mathfrak{R} = 0$. Two uncharged states under the Cartan $\mathfrak{h}\subset \mathfrak{su}(3)$ lead to two flavor modular differential equations, while the Sugawara condition gives an additional equation.

It is straightforward to verify that $|\mathcal{N}_T\rangle$ is an affine descendant of $|\mathcal{N}^\text{hw}_2\rangle$. At conformal-weight-three there are no affine null states, but there are descendants of null $\mathcal{N}^\text{hw}_2$ given by $J_1^A |\mathcal{N}_T\rangle$. In particular, the two null descendants $h^i_1|\mathcal{N}_T\rangle$ give rise to the two weight-three flavored modular differential equations. Similarly, $h^i_1 h^j_1|\mathcal{N}_T\rangle$ give three weight-two equations.

The algebra has four admissible modules with the highest weights
\begin{align}
	- \frac{3}{2} \widehat{\omega}_0, \qquad
	- \frac{3}{2} \widehat{\omega}_1, \qquad
	- \frac{3}{2} \widehat{\omega}_2, \qquad
	- \frac{1}{2} (\widehat{\omega}_0 + \widehat{\omega}_1 + \widehat{\omega}_2) \ .
\end{align}
The corresponding characters as $q$-series can also be solved from the equations from the (\ref{eqWeight4}), (\ref{eq:e43}) and (\ref{eq:e42}) \cite{Peelaers}. Their closed form are given by \cite{Kac:1988qc,2016arXiv161207423K}
\begin{align}
	\operatorname{ch}_{0} = & \ y^{-3/2} \frac{\eta(\tau)}{\eta(2 \tau)} \frac{
		\vartheta_1(\mathfrak{b}_1 - 2 \mathfrak{b}_2 |2 \tau)
		\vartheta_1( - \mathfrak{b}_1 - \mathfrak{b}_2 |2 \tau)
		\vartheta_1(-2 \mathfrak{b}_1 + \mathfrak{b}_2 |2 \tau)
	}{
		\vartheta_1(\mathfrak{b}_1 - 2 \mathfrak{b}_2 | \tau)
		\vartheta_1( - \mathfrak{b}_1 - \mathfrak{b}_2 | \tau)
		\vartheta_1(-2 \mathfrak{b}_1 + \mathfrak{b}_2 | \tau)
	} \ ,\\
	\operatorname{ch}_{1} = & \ - y^{-3/2} \frac{\eta(\tau)}{\eta(2 \tau)} \frac{
		\vartheta_4(\mathfrak{b}_1 - 2 \mathfrak{b}_2 |2 \tau)
		\vartheta_4( - \mathfrak{b}_1 - \mathfrak{b}_2 |2 \tau)
		\vartheta_1(-2 \mathfrak{b}_1 + \mathfrak{b}_2 |2 \tau)
	}{
		\vartheta_1(\mathfrak{b}_1 - 2 \mathfrak{b}_2 | \tau)
		\vartheta_1( - \mathfrak{b}_1 - \mathfrak{b}_2 | \tau)
		\vartheta_1(-2 \mathfrak{b}_1 + \mathfrak{b}_2 | \tau)
	} \ , \\
	\operatorname{ch}_{2} = & \ y^{-3/2} \frac{\eta(\tau)}{\eta(2 \tau)} \frac{
		\vartheta_1(\mathfrak{b}_1 - 2 \mathfrak{b}_2 |2 \tau)
		\vartheta_1( - \mathfrak{b}_1 - \mathfrak{b}_2 |2 \tau)
		\vartheta_1(-2 \mathfrak{b}_1 + \mathfrak{b}_2 |2 \tau)
	}{
		\vartheta_1(\mathfrak{b}_1 - 2 \mathfrak{b}_2 | \tau)
		\vartheta_1( - \mathfrak{b}_1 - \mathfrak{b}_2 | \tau)
		\vartheta_1(-2 \mathfrak{b}_1 + \mathfrak{b}_2 | \tau)
	} \ , \\
	\operatorname{ch}_3 = & \ y^{-3/2} \frac{\eta(\tau)}{\eta(2\tau)} \frac{
		\vartheta_4(\mathfrak{b}_1 - 2 \mathfrak{b}_2|2\tau)
	  \vartheta_4(-\mathfrak{b}_1 - \mathfrak{b}_2|2\tau)
	  \vartheta_4(-2\mathfrak{b}_1 + \mathfrak{b}_2|2\tau)
	}{
	  \vartheta_1(- \mathfrak{b}_1 - \mathfrak{b}_2|2\tau)
	  \vartheta_1(\mathfrak{b}_1 - 2 \mathfrak{b}_2|2\tau)
	  \vartheta_1(-2\mathfrak{b}_1 + \mathfrak{b}_2|2\tau)
	} \ .
\end{align}
It is straightforward to see that applying $T_{\mathbf{n}^\vee}$ with $\mathbf{n}^\vee = \alpha_1^\vee, \alpha_2^\vee, \alpha_1^\vee + \alpha_1^\vee$ generates all the highest weight characters,
\begin{align}
	& \ \operatorname{ch}_0(\mathfrak{b}_1 + \tau, \tau, \mathfrak{y} + K^{i1} \mathfrak{b}_i + \frac{1}{2}\tau) = \operatorname{ch_1} \ ,
	\quad
	\operatorname{ch}_0(\mathfrak{b}_2 + \tau, \tau, \mathfrak{y} + K^{i2} \mathfrak{b}_i + \frac{1}{2}\tau) = \operatorname{ch_2} \ , \nonumber\\
	& \ \operatorname{ch}_0(\mathfrak{b}_1 + \tau, \mathfrak{b}_2 + \tau, \tau, \mathfrak{y} + \mathfrak{b}_1 + \mathfrak{b}_2 + \tau) = \operatorname{ch}_3 \ .
\end{align}

\subsection{\texorpdfstring{$\widehat{\mathfrak{so}}(8)_{-2}$}{}}

The algebra $\widehat{\mathfrak{so}}(8)_{-2}$ is non-admissible. It is the associated chiral algebra of the 4d $\mathcal{N} = 2$ $SU(2)$ gauge theory with four fundamental flavors. The vacuum character, which is the Schur index of the $SU(2)$ gauge theory, has been computed analytically \cite{Pan:2021mrw}\footnote{See also \cite{Bourdier:2015wda,Bourdier:2015sga,Beemetal,Hatsuda:2022xdv} for analytic computations of many Schur indices.},
\begin{align}
	\operatorname{ch}_0
	= & \ \sum_{j = 1}^4 E_1\left[
    \begin{matrix}
      - 1 \\ m_j
    \end{matrix}
    \right]
    \frac{ i \vartheta_1(2\mathfrak{m}_j) }{\eta(\tau)}\prod_{\ell \ne j}\frac{\eta(\tau)}{\vartheta_1(\mathfrak{m}_j + \mathfrak{m}_\ell)}
    \frac{ \eta(\tau)}{\vartheta_1(\mathfrak{m}_j - \mathfrak{m}_\ell)} \nonumber \\
  = & \ \frac{\eta(\tau)^2}{\prod_{j = 1}^{4}\vartheta_1(2 \tilde{\mathfrak{b}}_j)}
  \sum_{\vec \alpha = \pm} \left(\prod_{i = 1}^{4}\alpha_i\right) E_2 \begin{bmatrix}
  	1 \\ \prod_{j = 1}^{4}\tilde b_j^{\alpha_j}  
	\end{bmatrix} \ .
\end{align}
where $\mathfrak{m}_{1,2} = \tilde {\mathfrak{b}}_1 \pm \tilde {\mathfrak{b}}_2$, $\mathfrak{m}_{3,4} = \tilde{\mathfrak{b}}_3 \pm \tilde{\mathfrak{b}}_4$, and the flavor fugacities $\mathfrak{m}_j$ are related to the ones in this paper by $\mathfrak{m}_1 = \mathfrak{b}_1, \mathfrak{m}_2 = \mathfrak{b}_2 - \mathfrak{b}_1$, $\mathfrak{m}_3 = - \mathfrak{b}_2 + \mathfrak{b}_3 + \mathfrak{b}_4$, $\mathfrak{m}_4 = \mathfrak{b}_3 - \mathfrak{b}_4$.

The flavored modular differential equations of the Schur index $\operatorname{ch}_0$ were studied in \cite{Peelaers,Zheng:2022zkm}, where there were $9+1$ equations at weight-two, $4$ equations at weight-three and one equation at weight-four. For $\mathfrak{so}(8)$, the representation $\mathfrak{R}$ is decomposed into three $\mathbf{35}$, each comes with three charge-zero states, and therefore account for the $9$ weight-two equations. These equations were shown to have four additional non-logarithmic solutions given by \cite{Peelaers,Zheng:2022zkm}
\begin{align}
	R_j \coloneqq
	\frac{i}{2}\frac{\vartheta_1(2\mathfrak{m}_j) }{\eta(\tau)}\prod_{\ell \ne j}\frac{\eta(\tau)}{\vartheta_1(\mathfrak{m}_j + \mathfrak{m}_\ell)}
	\frac{ \eta(\tau)}{\vartheta_1(\mathfrak{m}_j - \mathfrak{m}_\ell)} \ ,
\end{align}
which are the residues of the integrand that computes $\operatorname{ch}_0$. These solutions are linear combinations of the characters of the vacuum and other highest weight modules studied in \cite{Arakawa:2015jya}. Explicitly, the irreducible characters are given by spectral flows of the vacuum character, and they are related to the residues $R_j$ by \cite{2023arXiv230409681L}
\begin{align}
	\operatorname{ch}_{-2\widehat \omega_1} = & \ \operatorname{ch}_0 - 2R_1\\
	\operatorname{ch}_{-\widehat \omega_2} = & \ -2 \operatorname{ch}_0 + 2R_1 + 2R_2\\
	\operatorname{ch}_{-2\widehat \omega_3} = & \ \operatorname{ch}_0 - R_1 - R_2 - R_3 - R_4\\
	\operatorname{ch}_{-2\widehat \omega_4} = & \ \operatorname{ch}_0 - R_1 - R_2 - R_3 + R_4 \ .
\end{align}
In terms of the shift $T_{\vec n^\vee}$ in (\ref{T-operator}), we have\footnote{This is similar to the spectral flow discussed in \cite{2023arXiv230409681L}. The character of a twisted module is also accessible through spectral flow by fractional unit.}
\begin{align}
	(T_{- 2 \omega_i} + 1)\operatorname{ch}_0(\mathfrak{b}, \tau, \mathfrak{y}) = 2 \operatorname{ch}_{- 2 \widehat \omega_i} , \quad 
	(T_{- \omega_2} + 1)\operatorname{ch}_0(\mathfrak{b}, \tau, \mathfrak{y}) = - \operatorname{ch}_{- \widehat \omega_2} \ ,
\end{align}
where we simply replace $\mathbf{n}^\vee$ by $-2 \omega_{i = 1,3,4}$ and $- \omega_2$ (in $\alpha_i^\vee$ basis).

\subsection{\texorpdfstring{$(\widehat{\mathfrak{e}}_6)_{-3}$}{}}

The vacuum character of  $(\widehat{\mathfrak{e}}_6)_{-3}$ is identified with the Schur index $\mathcal{I}_{E_6}$ of the $E_6$ Minahan-Nemeschansky theory. The $E_6$ theory participates in a Argyres-Seiberg duality which relates its Schur index with that of the $SU(3)$ gauge theory with six fundamental hypermultiplets, which allows $\mathcal{I}_{E_6}$ to be written in terms of $\mathcal{I}_{SU(3)}$ \cite{Gadde:2010te,Razamat:2012uv,Pan:2021mrw},
\begin{align}\label{IE6}
  & \ \mathcal{I}_{E_6}(\vec c^{(1)}, \vec c^{(2)}, (wr, w^{-1}r, r^{-2})) \\
  = & \ \frac{\mathcal{I}_{SU(3) \ \text{SQCD}}(\vec c^{(1)}, \vec c^{(2)}, \frac{w^{\frac{1}{3}}}{r}, \frac{w^{- \frac{1}{3}}}{r})_{w \to q^{\frac{1}{2}}w}}{\theta(w^2)} + \frac{\mathcal{I}_{SU(3) \ \text{SQCD}}(\vec c^{(1)}, \vec c^{(2)}, \frac{w^{\frac{1}{3}}}{r}, \frac{w^{- \frac{1}{3}}}{r})_{w \to q^{ - \frac{1}{2}}w}}{\theta(w^{ - 2})} \ , \nonumber
\end{align}
where the elliptic theta function $\theta$ is related to the Jacobi theta function $\vartheta_1$ by
\begin{equation}
	\theta(z) \equiv \frac{\vartheta_1( \mathfrak{z})}{i z^{-\frac{1}{2}} q^{\frac{1}{8}} (q;q)} \ .
\end{equation}
Note that the Schur index $\mathcal{I}_\text{SQCD}$ and therefore $\mathcal{I}_{E_6}$ have closed form expressions in terms of $\vartheta$ and Eisenstein series \cite{Pan:2021mrw}. The index $\mathcal{I}_{SU(3)}$ is written in the form $\mathcal{I}_{SU(3)}(\vec c^{(1)}, \vec c^{(2)}, d^{(1)}, d^{(2)})$ where $\vec c^{(i)}$ denotes two copies of $SU(3)$ flavor fugacities and $d^{(i)}$ denotes two copies of $U(1)$ flavor fugacities. The $SU(3)^3 \subset E_6$ flavor fugacities are then given by $c^{(1, 2)}$, and $\vec c^{(3)} = (wr, w^{-1} r, r^{-2})$. The fugacities $c^{(1,2,3)}$ can be further reorganized into the $b$-variables with respect to $E_6$,
\begin{align}
	r = \sqrt{\frac{b_6}{b_3^{1/3}}}, \ w = \sqrt{\frac{b_6}{b_3}}, \ 
	c^{(1)}_1 = \frac{b_3^{1/3}}{b_1}, \ c^{(1)}_2 = \frac{b_2}{b_3^{2/3}}, \ 
	c^{(2)}_1 = \frac{b_4}{b_3^{2/3}}, \ c^{(2)}_2 = \frac{b_5}{b_3^{1/3}} \ , \nonumber
\end{align}
and one has the Schur index written as $\mathcal{I}_{E_6}(\mathfrak{b}, \tau, \mathfrak{y})$. At this stage, applying shifts $T_{\mathbf{n}^\vee}$ will generate new solutions to all the flavored modular differential equations, and therefore we believe that all highest weight characters of $(\widehat{\mathfrak{e}}_6)_{-3}$ can be obtained through suitable shift of the closed-form index (\ref{IE6}).


\section*{Acknowledgments}
The authors would like to thank Wolfger Peelaers for sharing Mathematica notebooks and many discussions. We also thank Bohan Li, Satoshi Nawata, Wenbin Yan, Jiahao Zheng for helpful discussions. Y.P. is supported by the National Natural Science Foundation of China (NSFC) under Grant No. 11905301.

\bibliographystyle{utphys2}

\bibliography{ref}

\end{document}